\documentclass[12pt]{article}
\usepackage{amsfonts,amsmath}
\usepackage{amssymb}

\makeatletter \@addtoreset{equation}{section} \makeatother

\DeclareMathOperator\arctanh{arctanh}

{\vspace{3mm} }

\def\al{\alpha}

\def\*{\star}

\def\E2{\mathbf{E}}
\def\w{\mathbf{w}}

\def\C{\mathcal C}
\def\T{\mathcal T}
\def\J{\mathbb J}

\newcommand{\be}{\begin{equation}}
\newcommand{\ee}{\end{equation}}
\newcommand{\bee}{\begin{eqnarray}}
\newcommand{\beee}{\begin{array}}
\newcommand{\eee}{\end{eqnarray}}
\newcommand{\eeee}{\end{array}}
\newcommand{\gb}{\beta}
\newcommand{\gga}{\gamma}

\newcommand{\gd}{\delta}

\newcommand{\gk}{\varkappa}
\newcommand{\gep}{\epsilon}

\newcommand{\gs}{\sigma}
\newcommand{\go}{\omega}

\newcommand{\dal}{\dot \alpha}
\newcommand{\dgb}{\dot \beta}
\newcommand{\dgga}{\dot \gamma}

\newcommand{\nn}{\nonumber}
\newcommand{\half}{\frac{1}{2}}

\newcommand{\p}{\partial}

\newcommand{\ff}{\frac}
\newcommand{\dr}{{\rm d}}

\begin{document}
\begin{flushright}
\end{flushright}

\vspace{0.5cm}
\begin{center}
{\large\bf Planar solutions of higher-spin theory. Nonlinear
corrections}

\vspace{1 cm}

\textbf{V.E.~Didenko and A.V.~Korybut}\\

\vspace{1 cm}

\textbf{}\textbf{}\\
\vspace{0.5cm}  \textit{I.E. Tamm Department of Theoretical
Physics,
Lebedev Physical Institute,}\\
 \textit{ Leninsky prospect 53, 119991, Moscow, Russia }\\
\par
\vspace{1 cm} didenko@lpi.ru,  akoribut@gmail.com
\end{center}

\vspace{0.4cm}

\begin{abstract}
\noindent Leading order higher-spin corrections to the linearized
higher-spin black brane are analyzed in four dimensions. It is
shown that the static solution that respects planar symmetry
exists in the bosonic case at given order. Its higher-spin Weyl
tensors are found in a closed form and are shown to have the
double copy origin. The effect of higher-spin fields to form a
strictly positive scalar condensate for any values of higher-spin
charges is observed.
\end{abstract}
\begin{flushright}
{\emph{In memory of Michael Soloviev}}
\end{flushright}

\newpage
\tableofcontents

\section{Introduction}\label{intro}

This paper is a continuation of the $AdS/CFT$
\cite{Maldacena:1997re}-\cite{Witten:1998qj} inspired study on
higher-spin (HS) analogs of GR black branes. For a detailed
motivation, extended introduction and references we refer to its
first part \cite{Didenko:2021vui}, where a linearized step in this
direction has been made and a closed form solution for the free HS
gauge fields proposed. The natural question we address here is
whether the HS brane like solution extends beyond free level and
if yes how does it look like? To this end we consider HS equations
of motion up to quadratic order.

The potential physical importance of this solution rests on the
experience in $AdS/CFT$, where black branes of (super)gravity play
a fundamental role as bulk dual states of thermal CFTs
\cite{arXiv:9808032}. The dictionary is achieved via an $AdS$
factor in the near horizon limit of the metric provided its
Euclidian time points are properly identified to make it smooth,
thus introducing a thermal $S^1$. In HS theory the line element
associated with $s=2$ metric is no longer gauge invariant object.
This seemingly revokes the above argument. However, since the
notion of HS global symmetries as transformations that leave a
particular solution invariant is similar to those in GR, it is
tempting to speculate that solutions with planar symmetry of HS
theory might be as important as they are in gravity from $AdS/CFT$
point of view.

Our main findings are the following. By analyzing the deformation
problem for planar solutions by interactions we come across with a
certain integrability condition that turns out to be satisfied for
the bosonic HS $A$--model. This allows us to find closed form
expressions for HS Weyl tensors. These reveal the double copy
structure and can be reconstructed in terms of multiple copies of
Maxwell tensor and its scalar zeroth copy. A particular attention
is paid to the linearized pure $s=2$ case corresponding to the
standard GR black brane. While the linearized solution survives in
the full GR, this is no longer so for HS interactions as spin two
sector receives corrections. Moreover, HS interactions induce non
zero $s=0$ and $s=4$ fields at leading order as well. We have also
studied the effect of HS fields to the scalar sector of the theory
and observed that they tend to induce a strictly positive scalar
condensate irrespectively of the signs of HS charges.

Interacting HS framework is well elaborated at the level of the
unfolded equations in four dimensions (see e.g.,
\cite{Vasiliev:1999ba}). A schematic form of these equations is
\begin{align}
&\dr\w+\w*\w=\Upsilon(\w, \w, C)+\Upsilon(\w, \w, C, C)+\dots\,,\label{hs1}\\
&\dr C+[\w, C]_*=\Upsilon(\w, C, C)+\Upsilon(\w, C, C,
C)+\dots\label{hs2}\,,
\end{align}
where $\w$ is the 1-form that encodes HS potentials, while $C$ is
the 0-form that contains information about HS Weyl tensors. Fields
$\w$ and $C$ both may appear in several copies like in e.g.,
supersymmetric models. $\Upsilon$ collectively denotes interaction
vertices. Formally, these can be extracted up to field
redefinition order by order from the generating Vasiliev system
\cite{more} which guarantees HS gauge invariance of vertices.
However, the gauge invariance alone is not sufficient for having
well-defined observables such as {\it e.g., } boundary correlation
functions. This further restricts functional class of the Vasiliev
master fields leading to the conjecture of spin locality
\cite{Gelfond:2018vmi}, \cite{Didenko:2018fgx}, which if true
guarantees finiteness of classical theory. Whether spin locality
applies to all orders is an open problem. Let us stress in this
respect that there are alternative approaches to HS problem in the
literature \cite{Aharony:2020omh}, where the locality is
sacrificed already at cubic order, while the guiding principle is
the bi-local approach of \cite{Das:2003vw} that by construction
gives proper boundary correlators at any order. The emergent
space-time and HS fields dual to a thermal $CFT$ was also studied
in \cite{Jevicki:2015sla} in bi-local terms.

There are few exact solutions of the Vasiliev equations found over
the years \cite{Prokushkin:1998bq}-\cite{Aros:2017ror}. We however
would like to analyze HS equations that follow from the generating
system in perturbations rather than the Vasiliev system itself in
this paper as this gives us control over locality of HS
interactions at lowest orders which otherwise is hard to trace
back. The detailed structure of these equations at leading order
will be given below. For now we note that the manifest form of
$\Upsilon$--vertices consistent with the locality criterium is
highly nontrivial and by now is known to few orders only.
Moreover, higher order locality itself seems so far problematic
\cite{Bekaert:2015tva}, \cite{Sleight:2017pcz} being a subject of
active research \cite{Vasiliev:2016xui}-\cite{Gelfond:2021two}.

Of our interest are leading order interactions in the sector of HS
Weyl tensors \eqref{hs2} terminated by local $\Upsilon(\w, C, C)$.
The corresponding perturbative expansion that we are going to work
out is the following
\begin{align}
\w=\w^{(0)}+\w^{(1)}\,,\\
C=C^{(1)}+C^{(2)}\,,
\end{align}
where $\w^{(0)}$ stands for the $AdS_4$ vacuum, while the
linearized contribution $\w^{(1)}$ and $C^{(1)}$ have been found
in \cite{Didenko:2021vui}. So, the equation we examine here is
\be\label{hs22}
\dr C^{(2)}+[\w^{(0)}, C^{(2)}]_*+[\w^{(1)},
C^{(1)}]_*=\Upsilon(\w^{(0)}, C^{(1)}, C^{(1)})\,.
\ee
Its static, planar solution for $C^{(2)}$, if exists, contains
second order corrections to HS black brane like Weyl tensors that
we would like to find and analyze.

The linearized solution \cite{Didenko:2021vui} we are going to use
for $\w^{(1)}$ and $C^{(1)}$ corresponds to HS planar excitations
of all integer spins each having its own 'mass like' real
parameter $m_s$. The reason we refer to it as to a HS analog of a
black brane rests on its space-time symmetry and its classical
double copy form rather than on any physical characteristics like
the presence of a horizon, which is by far beyond the reach in HS
theory. Nevertheless, many black hole solutions admit HS
generalization by forming a multiple copy of a certain spin $s=1$
field
\be\label{DCW}
\textnormal{Weyl}_s\sim (\textnormal{Maxwell})^s\,,
\ee
which in four dimensions says that the (anti)selfdual part of the
linearized spin $s$ Weyl tensor that appears on the left hand side
of this correspondence is given by $s$ copies of (anti)self-dual
Maxwell tensor. Relation \eqref{DCW} gets exact upon putting in a
proper dependence on a massless scalar field $\phi$ instead of
$\sim$, which itself turns out to be made of $s=1$ Maxwell tensor.
The case $s=2$ reproduces this way Petrov $D$ type Weyl tensor
that includes black hole solutions on $AdS$. In terms of gauge
potentials the multiple copies are realized via the Kerr-Schild
ansatz
\be\label{DCK}
\phi_{\mu_1\dots\mu_s}=\phi k_{\mu_1}\dots k_{\mu_s}\,,
\ee
which satisfies massless spin $s$ equation on $AdS$, provided
$k_{\mu}$ is the so-called Kerr-Schild vector. It makes Einstein
equations linearly exact, in particular. Eq. \eqref{DCW} for $s=2$
is called the Weyl double copy while its higher-spin extension is
naturally calling the Weyl multi-copy. Similarly, \eqref{DCK} is
the Kerr-Schild double copy for $s=2$ with the multi-copy
extension for any $s$. Properties \eqref{DCW} and \eqref{DCK} both
were originally observed within the unfolded approach in
\cite{Didenko:2008va}, \cite{Didenko:2009tc},
\cite{Didenko:2009td} for all four dimensional $AdS$ black holes
(including black branes) and then for the $AdS$ Kerr-Schild ones
in five dimensions\footnote{In the double copy literature
properties \eqref{DCK} and \eqref{DCW} for asymptotically
Minkowski black holes $(s=2)$ were independently observed in
\cite{Monteiro:2014cda}, \cite{Luna:2018dpt}.}
\cite{Didenko:2011ir}. The classical double copy origin of
Kerr-Schild black holes rests on their algebraically special form,
namely the $D$ Petrov type and thus may not be directly related to
the original KLT relations \cite{Kawai:1985xq}. Nevertheless there
seems to be a substantial recent progress in understanding the
classical double copy beyond algebraically special solutions
\cite{Adamo:2020qru}-\cite{Adamo:2021dfg}.

For our means we will greatly exploit the type $D$ structure of GR
black holes and in particular the linearized HS black brane
solution. It was shown that the Weyl double copy naturally results
from a general $AdS$ global symmetry parameter in
\cite{Didenko:2009tc} via the Penrose like transform\footnote{From
the twistor stand point this parameter literally corresponds to a
rank two dual twistor on asymptotically Minkowski space, see
\cite{Chacon:2021wbr}, \cite{White:2020sfn}.}
\cite{Didenko:2009td}, \cite{Didenko:2021vui}. Technically this
allows us plugging that parameter in the HS equations of motion
serving as a pillar for searching for their solutions with the
required symmetry. This gives us on the one hand a covariant
approach with respect to planar symmetry and on the other
considerably simplifies the analysis. In doing so we observe that
the parameter corresponding to the planar case induces a flat
connection \cite{Didenko:2021vui} that builds in into HS equations
making planar symmetry manifest.

We find the so called {\it current ansatz} very useful for
analysis of interactions. Originally introduced in
\cite{Vasiliev:2012vf} for establishing HS bulk-boundary
correspondence between fields and currents at first order, it
turns out to be applicable at second order too
\cite{Didenko:2017lsn} leading to a non-linear equation for HS
currents. Remarkably, the linearized planar solutions of
\cite{Didenko:2021vui} correspond to very simple conserved
currents. Namely, a spin $s$ planar excitation is given by a power
$s$ monomial current. This makes finding HS planar corrections in
the current space natural. To this end we covariantize current
ansatz of \cite{Vasiliev:2012vf} using the $AdS$ global symmetry
parameter as a kind of a compensator field and analyze second
order HS equation of motion in terms of currents.

We use the $4d$ supersymmetric Vasiliev HS equations
\cite{Vasiliev:1999ba}. This gives us several options for
embeddings of the linearized solution. Most of the time we study
the usual bosonic embedding. Another natural option is the so
called {\it chiral embedding} that appears upon association of
mostly positive helicity part of the linearized solution with one
of the two HS master zero-forms from the supersymmetric
twisted-adjoint sector, while mostly negative with another one.
Quite interestingly, the integrability condition that makes
quadratic corrections a deformation of the linearized solution
does not hold for chiral embedding. This means that either static
planar solutions do not exist beyond free level in this case or
they do not follow from deformation. In the latter case they can
not support the double copy either.

The paper is organized as follows. In section 2 we briefly recall
the structure of Vasiliev unfolded HS equations and give the
explicit form of schematically written eq. \eqref{hs22}. In
section 3 HS equations are rewritten in terms of the so called
current ansatz in a covariant with respect to planar symmetry way.
In section 4 we find solution to those equations, provide details
of its derivation and list important properties. Concluding
remarks are collected in the last section. Paper is supplemented
with two appendices one of which contains explicit check of the
static consistency condition, while another one includes
expression for function $h$ introduced in section 4.3.

\section{Higher-spin equations}
HS interactions \eqref{hs1}, \eqref{hs2} can be extracted from the
Vasiliev generating system \cite{Vasiliev:1999ba}. The procedure
is perturbative and starts with some exact solution.

\paragraph{Vacuum.}

The simplest HS vacuum state satisfies
\be\label{vacgo}
\dr\w+\w*\w=0\,,
\ee
\be\label{vacC}
C=0\,.
\ee
where $\w=\w(y,\bar y|x)$ is a 1-form that depends on the
auxiliary $sp(4)$ spinor variable $Y_{A}=(y_{\al}, \bar
y_{\dal})$. Decomposition of $\w$ in terms of $y$'s describes HS
gauge field sector. Non vacuum $C=C(y,\bar y|x)$ depends on $y$'s
too encoding gauge invariant physical degrees of freedom
propagating on some vacuum state $\w^{(0)}$. Star-product $*$ is
an associative product
\be\label{exp}
f(Y)*g(Y)=f(Y)e^{i\gep^{AB}\overleftarrow{\p}_A\overrightarrow{\p}_B}
g(Y)\,,
\ee
that defines HS algebra as the enveloping one of the following
oscillator relations
\be\label{com}
[y_\al, y_\gb]_*=2i\gep_{\al\gb}\,,\qquad [y_\al, \bar
y_{\dgb}]_*=0\,,\qquad [\bar y_{\dal}, \bar
y_{\dgb}]_*=2i\gep_{\dal\dgb}\,.
\ee
Purely gravitational solution of \eqref{vacgo} describes $AdS_4$
and is given by
\be\label{vac}
\w^{(0)}=-\ff{i}{4}(\go^{\al\gb}y_{\al}y_{\gb}+\bar{\go}^{\dal\dgb}\bar{y}_{\dal}\bar{y}_{\dgb}+
2{\bf e}^{\al\dgb}y_{\al}\bar{y}_{\dgb})\,,
\ee
where Lorentz connection $\go^{\al\gb}$, $\bar\go^{\dal\dgb}$ and
vierbein ${\bf e}^{\al\dgb}$ satisfy Cartan structure
equations\footnote{We set negative cosmological constant to a
number for convenience} of $AdS$
\begin{align}\label{adsOmega}
&\dr\go_{\al\gb}-\go_{\al}{}^{\gga}\go_{\gga\gb}-{\bf
e}_{\al}{}^{\dgga}{\bf e}_{\gb\dgga}=0\,,\\ \label{adsTetrad} &\dr
{\bf e}_{\al\dal}-\go_{\al}{}^{\gb}{\bf
e}_{\gb\dal}-\bar{\go}_{\dal}{}^{\dgb}{\bf e}_{\al\dgb}=0\,.
\end{align}

\paragraph{Free equations.} Free propagation of HS fields around
$AdS$ vacuum \eqref{vac} is described by the gauge invariant
twisted-adjoint covariant constancy condition on the 0-form
$\C(y,\bar y; k, \bar k |x)$ and by the on-mass-shell condition in
the sector of gauge fields parameterized by 1-form $\w(y, \bar y;
k, \bar k|x)$. Here, $k$ and $\bar k$ are the two Clifford
elements that make HS system $\mathcal{N}=2$ supersymmetric
satisfy
\begin{align}
&\{k,y_{\al}\}=0\,,\quad [k, \bar y_{\dal}]=0\,,\quad k^2=1\,,\label{k1}\\
&\{\bar k,\bar y_{\dal}\}=0\,,\quad [\bar k, y_{\al}]=0\,,\quad \bar k^2=1\,,\label{k2}\\
&[k, \bar k]=0\,.\label{k3}
\end{align}
Dynamical master fields are singled out by the following condition
\be\label{dyn}
\w(y, \bar y; -k, -\bar k|x)=\w(y, \bar y; k, \bar k|x)\,,\qquad
\C(y,\bar y; -k, -\bar k |x)=-\C(y,\bar y; k, \bar k |x)\,.
\ee
Those that do not satisfy \eqref{dyn} are the so called
topological fields \cite{Vasiliev:1999ba}, which describe no
propagating degrees of freedom and we set them to zero in what
follows meaning that
\be\label{C}
\C=C(y,\bar y|x)k+\bar C(y,\bar y|x)\bar k\,,
\ee
\be\label{w}
\w=\go(y,\bar y|x)+\go'(y,\bar y|x)k\bar k\,.
\ee
Proper reality conditions on master fields should be further
imposed
\be
\w^{\dagger}=-\w\,,\qquad \C^{\dagger}=\C\,,
\ee
where
\be
y^{\dagger}_{\al}=\bar y_{\dal}\,,\qquad k^{\dagger}=\bar k\,,
\ee
implying
\begin{align}
&C^\dagger=\bar C(y, -\bar y)\,,\qquad \bar C^\dagger=C(-y, \bar
y)\,,\\
&\go^\dagger=-\go(y, \bar y)\,,\qquad \go'^{\dagger}=-\go'(-y,
-\bar y)\,.
\end{align}
For the purely bosonic spin-statistics which is our case of
interest one has in addition
\be
\C(y, -\bar y)=\C(-y, \bar y)\,,\qquad w(y, -\bar y)=w(-y, \bar
y)\,.
\ee
Free field dynamics is governed by the following equation of
motion \cite{Vasiliev:1988sa}
\be\label{OMS}
D_0 \w=\ff{i\bar \eta}{4}{\bf e}^{\al\dgga}{\bf
e}^{\gb}{}_{\dgga}\ff{\p^2}{\p y^{\al}\p y^{\gb}}\C(y,0|x)\bar k+
\ff{i\eta}{4}{\bf e}^{\gga\dal}{\bf e}_{\gga}{}^{\dgb}\ff{\p^2}{\p
\bar y^{\dal}\p \bar y^{\dgb}}\C(0,\bar y|x)k\,,
\ee
where
\be
D_{0}=\dr+\go^{\al\gb}y_{\al}\ff{\p}{\p
y^{\gb}}+\bar{\go}^{\dal\dgb}\bar{y}_{\dal}\ff{\p}{\p \bar
y^{\dgb}}+ {\bf e}^{\al\dgb}\left(y_{\al}\ff{\p}{\p \bar
y^{\dgb}}+\bar{y}_{\dgb}\ff{\p}{\p y^{\al}}\right)\label{D0}
\ee
is the $AdS_4$ covariant derivative and $\eta$ is an arbitrary
constant phase factor which generally breaks parity of the theory
beyond free level unless $\eta=1$ or $\eta=i$ (see
\cite{Sezgin:2003pt}). In this paper we consider parity even case
with $\eta=1$. Master field $\C$ is subject to the twisted-adjoint
constancy condition
\be\label{tw}
D \C-i{\bf e}^{\al\dal}\left(y_{\al}\bar y_{\dal}-\ff{\p}{\p
y^{\al}}\ff{\p}{\p\bar y^{\dal}}\right)\C=0\,,
\ee
where $D$ is the Lorentz covariant derivative
\be\label{lord}
D=\dr+\go^{\al\gb}y_{\al}\ff{\p}{\p
y^{\gb}}+\bar{\go}^{\dal\dgb}\bar{y}_{\dal}\ff{\p}{\p \bar
y^{\dgb}}\,.
\ee

\paragraph{Quadratic corrections} Local HS non-linear
corrections at leading order can be systematically extracted from
Vasiliev equations \cite{Vasiliev:2016xui}. The resulting
contribution to the right hand side of \eqref{tw} that contains
the minimal number of derivatives acquires the following
form\footnote{The vertex contains bosons and fermions as well as
parity breaking parameter $\eta$. Its bosonic reduction for
type--$A$ model with $\eta=1$ is known in the partially gauged
fixed form at the Lagrangian level \cite{Sleight:2016dba}.}
\begin{align}\label{CCeq}
&D \C-i{\bf e}^{\al\dal}\left(y_{\al}\bar y_{\dal}-\ff{\p}{\p
y^{\al}}\ff{\p}{\p\bar y^{\dal}}\right)\C=-[\w,\C]_*-\\
&-\ff{\eta}{2}{\bf e}^{\al\dal}y_{\al}\int\ff{\dr \bar u\dr\bar
v}{(2\pi)^2}\int_{0}^{1} \dr\tau
e^{i\bar{u}_{\dal}\bar{v}^{\dal}}\left((1-\tau)\p_{1}-\tau\p_{2}\right)_{\dal}\C(\tau
y,\bar y+\bar u)\C(-(1-\tau)y, \bar y+\bar v)k-\nn\\
&-\frac{\bar{\eta}}{2}\mathbf{e}^{\alpha \dot{\alpha}}
\bar{y}_{\dot{\alpha}} \int \frac{\dr u\, \dr v}{(2\pi)^2}
\int_0^1 \dr \tau \, e^{iu_\alpha
v^\alpha}\big((1-\tau)\partial_1-\tau \partial_2\big)_\alpha
\mathcal{C}(y+u,\tau
\bar{y})\mathcal{C}(y+v,(\tau-1)\bar{y})\bar{k}\,,\nn
\end{align}
where star-product $*$ is attributed to the HS algebra and
realized via the Moyal product \eqref{exp}
and $\p_{1\dal}$, $\p_{1\al}$ differentiate argument of the first
field (as seen from the left) of the two $\C\C$ in \eqref{CCeq},
while $\p_{2\dal}$, $\p_{2\al}$ act on the second $\C$
correspondingly. One should be cautious with Kleinians $k$ and
$\bar k$ when performing $\bar u, \bar v$ -- integration. General
rule is to drag all the Kleinians that may enter fields $\C's$ and
$\w$, say, to the right using \eqref{k1}-\eqref{k3} before
integration\footnote{Note that in doing so the (anti)holomorphic
Klein operator flips the sign of the whole (anti)holomorphic
argument of $\C$ or $w$ rather than $ (\bar y)\, y$ alone.}.

There are two natural embeddings of the bosonic fields one of each
spin into the $\mathcal{N}=2$ supersymmetric theory with two
fields of each spin.

\paragraph{Bosonic embedding} There is a well known truncation of
the supersymmetric theory that leads to the bosonic HS model
having each field appearing once. It is governed by certain
automorphism of the full Vasiliev equations, which allows one to
set
\begin{align}
&\C=C(y, \bar y|x)(k+\bar k)\,,\qquad \w=\go(y,\bar y|x)(1+k\bar k)\label{bosembd}\\
&C(y,-\bar y)=C(-y, \bar y)\,,\qquad \go(y, -\bar y)=\go(-y, \bar
y)\,,
\end{align}
where master fields $C$ and $\go$ describe single copy of each
integer spin. Substituting \eqref{bosembd} into \eqref{CCeq} one
arrives at the equation for quadratic HS correction for fields
$s=0, 1, 2, \dots$
\begin{align}\label{CCboseq}
&D C-i{\bf e}^{\al\dal}\left(y_{\al}\bar y_{\dal}-\ff{\p}{\p
y^{\al}}\ff{\p}{\p\bar y^{\dal}}\right)C=-\go*C+C*\pi(\go)-\\
&-{\eta}{\bf e}^{\al\dal}y_{\al}\int\ff{\dr \bar u\dr\bar
v}{(2\pi)^2}\int_{0}^{1} \dr\tau
e^{i\bar{u}_{\dal}\bar{v}^{\dal}}\left((1-\tau)\p_{1}-\tau\p_{2}\right)_{\dal}C(\tau
y,\bar y+\bar u)C((1-\tau)y, \bar y+\bar v)+\nn\\
&-{\bar{\eta}}{\bf e}^{\al\dal}\bar{y}_{\dot{\al}}\int\ff{\dr  u\,
\dr v}{(2\pi)^2}\int_{0}^{1} \dr\tau
e^{iu_{\al}v^{\al}}\left((1-\tau)\p_{1}-\tau\p_{2}\right)_{\al}C(y+u,\tau
\bar{y})C( y+v,(1-\tau)\bar{y}\,,\nn
\end{align}
where
\be
\pi f(y, \bar y)=f(-y, \bar y)\,,\qquad \bar\pi f(y, \bar y)=f(y,
-\bar y)\,.
\ee
For bosonic fields $\pi f=\bar\pi f$.

\paragraph{Chiral embedding} While the bosonic embedding is
defined at the level of full equations of motion, there is yet
another embedding capturing each spin $s$ field only once that can
be defined at least perturbatively in four dimensions. Indeed,
consider \eqref{C}. Since $C$ is complex it corresponds to two
real fields in general. We can however set in \eqref{C}
\be
C(y,\bar y|x)=C^{+}(y,\bar y|x)\,,\qquad \bar C(y,\bar
y|x)=C^{-}(y,\bar y|x)\,,
\ee
where $C^+$ is the helicity positive part and $C^{-}$ is the
helicity negative one of one and the same field $C^++C^-$
describing a real HS module. Since both $C^{\pm}$ satisfy
\eqref{tw} and
\be
\bar C^{\pm}=C^{\mp}
\ee
it follows that \eqref{C} and \eqref{w}
\be\label{Cwch}
\C=C^+ k+C^-\bar k\,,\qquad \w=\go (1+k\bar k)
\ee
describe each spin $s$ once, where we take for $s>0$
\be
C^-(0,\bar y)=0\,,\qquad C^+(y, 0)=0\,.
\ee

\section{Current ansatz}
We now proceed with the so called {\it current form} of HS
equations introduced in \cite{Vasiliev:2012vf} at free level
within the $AdS/CFT$ approach. It is designed to extract HS
current module from fields in the twisted-adjoint. By construction
it uses $3+1$ decomposition in terms of boundary coordinates $x^i$
and radial $z$. We want to do something similar but adjusted to
planar symmetry. We prefer to stay covariant however and for that
reason introduce a proper $AdS_4$ global symmetry parameter
following \cite{Didenko:2021vui}.

\subsection{Global symmetry parameter and a flat connection}
Planar solutions of the linear HS equations both in the
twisted-adjoint \eqref{tw} and adjoint \eqref{OMS} sectors have
been recently obtained using the generating $AdS$ global symmetry
parameter in \cite{Didenko:2021vui}. The planar symmetry also
provides one with an auxiliary flat connection which plays a
crucial role in solving gauge field sector. All necessary details
of that construction can be found in \cite{Didenko:2021vui}. Here
we briefly recall the necessary formulae, which we are going to
use in what follows.

The $AdS_4$ global symmetry parameter $K_{AB}=K_{BA}$
\be\label{param} K_{AB}=\left(
\begin{array}{cc}
\gk_{\al\gb} & v_{\al\dgb}\\
v_{\gb\dal} & \bar{\gk}_{\dal\dgb}\\
\end{array}
\right)\,,\quad \gk_{\al\gb}=\gk_{\gb\al}\,,\quad
\bar{\gk}_{\dal\dgb}=\bar{\gk}_{\dgb\dal}\,,
\ee
is known to generate $D$ -- type solutions\footnote{Minkowski
limit for $K_{AB}$ corresponds to what is known as rank two dual
twistor, that generates $D$ type solutions, see e.g.,
\cite{White:2020sfn}.} both in GR \cite{Didenko:2009tc} and in HS
theory \cite{Didenko:2009td}. It satisfies
\begin{align}
&D\gk_{\al\gb}={\bf e}_{\al}{}^{\dgga}v_{\gb\dgga}+{\bf e}_{\gb}{}^{\dgga}v_{\al\dgga}\,,\label{param2a}\\
&D v_{\al\dal}={\bf e}_{\al}{}^{\dgga}\bar{\gk}_{\dal\dgga}+ {\bf
e}^{\gga}{}_{\dal}\gk_{\gga\al}\,,\label{param2b}
\end{align}
where $v_{\al\dgb}$ is an $AdS_4$ Killing vector, while
$\gk_{\al\gb}$, $\bar\gk_{\dal\dgb}$ are the (anti)self-dual
components of the so called closed conformal Killing-Yano tensor
(see \cite{Didenko:2021vui} for more details). While different
$K$'s correspond to different type of solutions it generates, the
planar case is singled out by the following condition\footnote{We
change notation of \cite{Didenko:2021vui} by introducing $z=\ff
1r$. }
\be\label{brane}
K_{A}{}^{C}K_{CB}=0\,,\qquad
z^{-2}:=-\ff12\gk_{\al\gb}\gk^{\al\gb}=-\ff12\bar\gk_{\dal\dgb}\gk^{\al\gb}\geq
0
\ee
or more explicitly
\begin{align}
&\gk_{\al}{}^{\gga}\gk_{\gga\gb}+v_{\al}{}^{\dgga}v_{\gb\dgga}=0\,,\label{rel1}\\
&\gk_{\al}{}^{\gga}v_{\gga\dgb}=\bar{\gk}_{\dgb}{}^{\dgga}v_{\al\dgga}\,,\label{rel2}\\
&\bar\gk_{\dal}{}^{\dgga}\bar\gk_{\dgga\dgb}+v^{\gga}{}_{\dal}v_{\gga\dgb}=0\,.\label{rel3}
\end{align}
If \eqref{brane} is satisfied then one can show that an $sp(2)$
flat connection shows up
\be\label{post}
{\w}_{\al\gb}=\go_{\al\gb}-\ff1z\, {\bf E}_{\al\gb}\,,\qquad \bar
\w_{\dal\dgb}=\bar{\go}_{\dal\dgb}-\ff1z\, {\bf E}_{\dal\dgb}\,,
\ee
where by means of \eqref{rel1}-\eqref{rel3} one introduces {\it
metric}
\be\label{k}
k_{\al\dal}=-z^2\gk_{\al}{}^{\gb}v_{\gb\dal}\,,\qquad
k_{\al}{}^{\dgga}k_{\gb\dgga}=\gep_{\al\gb}\,,\qquad
k^{\gga}{}_{\dal}k_{\gga\dgb}=\gep_{\dal\dgb}
\ee
that allows one converting dotted indices into undotted ones and
vice versa according to the following convention. For any
$A_{\dal}$ we can define
\be\label{conv1}
A_{\al}:=k_{\al}{}^{\dgb}A_{\dgb}\,,
\ee
which entails from \eqref{k}
\be\label{conv2}
A_{\dal}=k^{\gga}{}_{\dal}A_{\gga}\,.
\ee
With this conventions one can define one-forms made of vierbein
${\bf e}_{\al\dgb}$
\be
{\bf E}_{\al\gb}=\ff{z}{2}({\bf e}_{\al\gb}+{\bf e}_{\gb\al})\,,
\ee
\be
{\bf E}=z\,{\bf e}_{\al}{}^{\al}\,.
\ee
The result of these definitions is the $sp(2)$ flatness condition
for \eqref{post}
\be
\dr \w_{\al\gb}-\w_{\al}{}^{\gga}\w_{\gb\gga}=0\,,\qquad \dr
\bar\w_{\dal\dgb}-\bar\w_{\dal}{}^{\dgga}\bar\w_{\dgb\dgga}=0\,.
\ee
It is natural then to define the following differential
\be\label{covdif}
\nabla A_{\al\dal}=\dr
A_{\al\dal}-\w_{\al}{}^{\gb}A_{\gb\dal}-\bar{\w}_{\dal}{}^{\dgb}A_{\al\dgb}\,,\qquad
\nabla^2=0\,,
\ee
which provides
\be\label{const0}
\nabla\left(z{\gk_{\al\gb}}\right)=
\nabla\left(z{v_{\al\dgb}}\right)=
\nabla\left(z{\bar\gk_{\dal\dgb}}\right)=\nabla k_{\al\dgb}=0
\ee
and
\begin{align}\label{d1r}
\nabla{\bf E_{\al\gb}}=\nabla {\bf E}=0\,,\qquad \dr z={\bf E}\,.
\end{align}
Let us note that the only Lorentz scalar one can construct for the
planar case out of components \eqref{param} is $z$, \eqref{brane}.
However, one can define a $\dr$--exact 1-form
\be\label{time}
\dr t=-\ff z2\gk_{\al\gb}{\bf E}^{\al\gb}=\ff12{\bf
e}^{\al\dal}v_{\al\dal}\,,
\ee
which has the meaning of time differential as we can see using the
Poincar\'{e} chart. It should be stressed though, that $t$ itself
can not be expressed via $K_{AB}$.

\paragraph{Poincar\'{e} realization}
The form of global symmetry parameter $K_{AB}$ \eqref{param} is
particularly simple in Poincar\'{e} coordinates $(t, x, y, z)$
\be\label{Poincare}
ds^2=\ff{1}{z^2}(-\dr t^2+\dr x^2+\dr y^2+\dr z^2)\,.
\ee
Let us choose vierbein
\be
{\bf e}^0=\ff{\dr t}{z}\,,\qquad {\bf e}^{1}=\ff{\dr
z}{z}\,,\qquad {\bf e}^{2}=\ff{\dr x}{z}\,,\qquad {\bf
e}^3=\ff{\dr y}{z}\,,
\ee
which in spinor terms is given by
\be
{\bf e}^{\al\dgb}=\ff1z\begin{pmatrix} {\dr t}+\dr y  & {\dr z}-i\,\dr x \\
{\dr z}+i\,\dr x & {\dr t}-\dr y
\end{pmatrix}\,.
\ee
Then  Killing vector
\be
v^{\mu}=(1,0,0,0)
\ee
generates the following Lorentz components of our global symmetry
parameter $K_{AB}$,  \cite{Didenko:2015pjo}
\be
v_{\al\dgb}=\ff1z \begin{pmatrix}1  & 0\\
0 & 1\end{pmatrix}\,,\qquad \gk_{\al\gb}=\ff1z \begin{pmatrix}1  & 0\\
0 & -1\end{pmatrix}\,,\qquad \bar\gk_{\dal\dgb}=\ff1z \begin{pmatrix}1  & 0\\
0 & -1\end{pmatrix}\,.
\ee
Written down in the Poincar\'{e} coordinates it has the form
\be
K_{AB}=\ff 1z\begin{pmatrix} 1 & 0 & 1 & 0\\
0 & -1 & 0 & 1\\
1 & 0 & 1 & 0\\
0 & 1 & 0 & -1
\end{pmatrix}
\ee
manifestly satisfying \eqref{brane}. Note, that definition of $z$
\eqref{brane} literally reproduces the radial Poincar\'{e}
coordinate, while \eqref{time} gives Poincar\'{e} time
differential $\dr t$. We are not going to use explicit coordinates
anywhere in our paper though leaving them here as a useful
illustration.

\subsection{$\nabla$-- covariant form of HS equations and the
current ansatz}

Let us now bring covariant differential \eqref{covdif} into the
twisted adjoint sector of HS equations \eqref{tw}, \eqref{CCeq}.
For that matter we should first rewrite it in terms of oscillator
action
\be\label{nabla}
\nabla=\dr+\w^{\al\gb}y_{\al}\ff{\p}{\p
y^{\gb}}-\bar{\w}^{\al\gb}\bar y_{\al}\ff{\p}{\p \bar
y^{\gb}}\,,\qquad \nabla^2=0\,,
\ee
where $\w_{\al\gb}$ and $\bar\w_{\al\gb}$ are given in
\eqref{post}. Using \eqref{lord} we then obtain
\be
D=\nabla+\ff1z{\bf E}^{\al\gb}\left(y_{\al}\ff{\p}{\p
y^{\gb}}-\bar y_{\al}\ff{\p}{\p\bar y^{\gb}}\right)\,.
\ee
Substituting this into \eqref{tw} gives
\be\label{tw1}
\nabla \C+\ff{i}{z} {\bf E}^{\al\gb}(y+i\bar \p)_{\al}(\bar
y-i\p)_{\gb}\C+\ff{i}{2z}{\bf E}(y_{\al}\bar
y^{\al}+\p_{\al}\bar\p^{\al})\C=0\,.
\ee

An important observation that makes problem of non-linear
corrections especially interesting is the HS Fock projector that
appears in the linearized solution uniformly for all spins
\cite{Didenko:2021vui}. This leads us to introduce the exponential
that satisfies the following projector condition
\be\label{Fock}
4e^{i y_{\al}\bar y^{\al}}*4e^{i y_{\al}\bar y^{\al}}=4e^{i
y_{\al}\bar y^{\al}}\,,
\ee
which in addition is covariantly constant
\be\label{fc}
\nabla e^{i y_{\al}\bar y^{\al}}=0\,.
\ee
Fock projector naturally appears in bulk-boundary analysis of HS
equations \cite{Vasiliev:2012vf} and as shown in
\cite{Didenko:2017lsn} decouples from second order equation
\eqref{CCeq} reducing an infinite dimensional twisted-adjoint
module to a certain current-type equation that admits polynomial
solutions. Given these facts it is natural to adapt HS equations
by making them covariant under flat connection \eqref{post} and
proceed along the lines of \cite{Vasiliev:2012vf},
\cite{Didenko:2017lsn} at higher orders by introducing the {\it
current ansatz}. To this end let us define a {\it current module}
$\T(w,\bar w; k, \bar k|x)$ via the following condition
\be\label{CT1}
\C=z\, e^{iy_{\al}\bar y^{\al}}\T(w,\bar w; k, \bar k|x)\,,
\ee
where
\be
w_{\al}={\sqrt z}y_{\al}\,,\qquad \bar w_{\al}={\sqrt z}\bar
y_{\al}\,.
\ee
Feeding \eqref{CT1} into \eqref{tw1} and using \eqref{fc} results
in the following linear equation for current $\T$
\be\label{Teq}
\nabla\T+i\, {\bf E}^{\al\gb}\,\p_{\al}\bar\p_{\gb}
\T+\ff{i}{2}{\bf E}\,\p_{\al}\bar\p^{\al}\T=0\,,
\ee
where $\p_{\al}$ and $\bar\p_{\al}$ are partial derivatives with
respect to $w$ and $\bar w$ correspondingly, while covariant
differential \eqref{nabla} is redefined to act properly on
functions of $w$ and $\bar w$
\be\label{nablaw}
\nabla:=\dr+\w^{\al\gb}w_{\al}\ff{\p}{\p
w^{\gb}}-\bar{\w}^{\al\gb}\bar w_{\al}\ff{\p}{\p \bar
w^{\gb}}\,,\qquad \dr w_{\al}=\dr\bar w_{\al}=0\,.
\ee

Remarkable property of ansatz \eqref{CT1} is that it goes through
nonlinear HS equation \eqref{CCeq} factoring out Fock exponential
$e^{iy_{\al}\bar y^{\al}}$ from the nonlinear right hand side.
This phenomenon takes place due to a somewhat specific form of the
interaction term in \eqref{CCeq}. In particular, one may note that
Fock projector \eqref{Fock} being rescaled by the homotopy
integration variable $\tau$ as prescribed by \eqref{CCeq} commutes
with operator $(1-\tau)\p_{1}-\tau\p_{2}$ that enters the
quadratic vertex. Substituting \eqref{CT1} into \eqref{CCeq} and
after some simple algebra that includes change of integration
variables one arrives at
\begin{align}\label{TT}
&\nabla\T+i\, {\bf E}^{\al\gb}\,\p_{\al}\bar\p_{\gb}
\T+\ff{i}{2}{\bf
E}\,\p_{\al}\bar\p^{\al}\T={\bf \J}^c+{\bf \J}^\go\,,\\
&\J^c[\T, \T]={\bf E}^{\al\gb}\mathcal{J}_{\al\gb}[\T,\T]+{\bf
E}\mathcal{J}[\T,\T]\,,
\end{align}
where $\J^c$ stems from the current interaction in the second line
of \eqref{CCeq}, while $\J^\go$ is the gauge part of commutator
$[\w, C]_*$. Explicitly, current contribution $\J^c$ has the form
\begin{align}\label{JCCaa}
&\mathcal{J}_{\al\gb}=-\frac{\eta}{4z^2}\,w_{\al}\int\ff{\dr \bar
u\dr\bar v}{(2\pi)^2}\int_{0}^{1} \dr\tau
e^{-\ff{i}{z}\bar{u}_{\al}\bar{v}^{\al}}\left((1-\tau)\p_{\bar
u}-\tau\p_{\bar v}\right)_{\gb}\times\nn\\
&\times\T(\tau w,\bar u+\bar w+(1-\tau)w)\T(-(1-\tau)w, \bar
v+\bar w-\tau w)k+(\al\leftrightarrow\gb)\nn+\\
&+\frac{\bar{\eta}}{4z^2}\,\bar w_{\al}\int\ff{\dr u\dr
v}{(2\pi)^2}\int_{0}^{1} \dr\tau
e^{\ff{i}{z}{u}_{\al}{v}^{\al}}\left((1-\tau)\p_{u}-\tau\p_{v}\right)_{\gb}\times\\
&\times\T(u+w+(1-\tau)\bar w,\tau\bar w)\T(v+w-\tau\bar
w,-(1-\tau)\bar w)\bar k-(\al\leftrightarrow\gb)\,,\nn
\end{align}
\begin{align}\label{JCC}
&\mathcal{J}=\frac{\eta}{4z^2}\,w^{\al}\int\ff{\dr \bar u\dr\bar
v}{(2\pi)^2}\int_{0}^{1} \dr\tau
e^{-\ff{i}{z}\bar{u}_{\al}\bar{v}^{\al}}\left((1-\tau)\p_{\bar
u}-\tau\p_{\bar v}\right)_{\al}\times\nn\\
&\times\T(\tau w,\bar u+\bar w+(1-\tau)w)\T(-(1-\tau)w, \bar
v+\bar w-\tau w)k\nn-\\
&+\frac{\bar{\eta}}{4z^2}\,\bar w^{\al}\int\ff{\dr u\dr
v}{(2\pi)^2}\int_{0}^{1} \dr\tau
e^{\ff{i}{z}{u}_{\al}{v}^{\al}}\left((1-\tau)\p_{u}-\tau\p_{v}\right)_{\al}\times\\
&\times\T(u+w+(1-\tau)\bar w,\tau\bar w)\T(v+w-\tau\bar
w,-(1-\tau)\bar w)\bar k\,,\nn
\end{align}
while gauge part $\J^\go$ is
\begin{align}\label{JwC}
&\J^\go=\ff{1}{(2\pi)^4 z^4}\int{\dr u\dr v \dr\bar u \dr\bar v}
e^{\ff{i}{z}(u_{\al}v^{\al}-\bar
u_{\al}\bar v^{\al})}\times\\
&\times\big(\T(w+u, \bar u+\bar w)\w(v-\bar u+w-\bar w, \bar
v+\bar w-w)-\\
&-\w(u+w+\bar w, \bar u+v+w+\bar w)\T(v+w, \bar w+\bar
v)\big)\nn\,.
\end{align}
We warn the reader on manipulations with Klein operators $k$ and
$\bar k$ that fields $\T$ and $\w$ implicitly contain in the above
formulae. When dragging, say, $k$ to the right through a master
field be it $\T$ or $\w$ one should change the sign of its whole
first spinorial argument. Similarly for $\bar k$ and the second
argument. In other words, one should use the following rule, e.g.,
$k\T(a,b)=\T(-a, b)k$.

\section{Solutions}
Recall that our goal is to solve \eqref{CCeq} which reads
schematically
\be
D \C^{(2)}-i{\bf e}^{\al\dal}\left(y_{\al}\bar y_{\dal}-\ff{\p}{\p
y^{\al}}\ff{\p}{\p\bar y^{\dal}}\right)\C^{(2)}=\bf{J}(\C^{(1)},
\C^{(1)})\,,
\ee
where $\C^{(1)}$ comes from the first order solution sourcing
$\C^{(2)}$ at the second. Note that $\bf{J}(\C,\C)$ contains
contribution from two pieces, the current-type on the second line
of \eqref{CCeq} and the gauge one $[\w, \C]$, which is also of
$\C\C$ -- type, since $\w=\w[\C]$. The source is made of
linearized solution \cite{Didenko:2021vui}. It is static and has
two-dimensional planar symmetry. We are interested in quadratic
corrections $\C^{(2)}$ that maintain this symmetry. To find these
we therefore need explicit form of $\bf{J}(\C^{(1)}, \C^{(1)})$.
It is much simpler however analyzing \eqref{TT} and for that the
free solution should be taken from \eqref{Teq}. Since the
linearized solution of \cite{Didenko:2021vui} was obtained for HS
{\it A--model} from now on we set
\be\label{A}
\eta=\bar\eta=1\,.
\ee

\subsection{Linear solutions} At free level the final result of
\cite{Didenko:2021vui} that solves \eqref{tw} and \eqref{OMS}
is\footnote{Note that the exponential projector factor in $C$ is
unique for any spin $s$. This behavior is typical of HS
bulk-to-boundary propagators \cite{GY1}, \cite{Didenko:2012vh}
that contain similar projector for every spin in the
twisted-adjoint.}
\begin{align}
&C^{(1)}=z(f(y)+f(\bar y))e^{i y_{\al}\bar
y^{\al}}\,,\label{Csol1}\\
&\w^{(1)}=-\ff{i}{2}{\bf
E}^{\al\gb}\int_{0}^{1}\dr\tau\p_{\al}\p_{\gb}f(\tau
y+(1-\tau)\bar y)\,,\label{wsol}
\end{align}
where
\be\label{feq}
f(y):=f\left(\ff{z^2}{2}\gk_{\al\gb}y^{\al}y^{\gb}\right)\,,\qquad
f(\bar y):=f\left(\ff{z^2}{2}\gk_{\al\gb}\bar y^{\al}\bar
y^{\gb}\right)
\ee
is an arbitrary function of its bilinear in $y$'s argument and
$\p_{\al}$ in \eqref{wsol} differentiates a spinorial argument.
Eqs. \eqref{Csol1} and \eqref{wsol} describe bosonic fields only.
For any $s>0$ the solution is static and possesses planar
symmetry\footnote{The case of $s=0$ is special since it has larger
global symmetry which is the boundary Poincar\'{e} algebra.}. For
example, $s=2$ component in \eqref{Csol1} corresponds to the Weyl
tensor of a black brane. The whole freedom of the solution is
stored in the Taylor coefficients of function $f(x)$, which are
associated with spin $s$ parameters. A particular example of
$f_s(x)$ that describes a single spin $s$ field is
\be\label{feqs}
f_s(y)=\ff{m_s}{s!}
\left(\ff{z^2}{2}\gk_{\al\gb}y^{\al}y^{\gb}\right)^s\,,
\ee
where $m_s$ is an arbitrary real number. In fact, $f$ plays a role
of a generating function for the primary HS Weyl tensors of the
form
\be\label{Dtype}
C^{(1)}_{\al_1\dots\al_{2s}}={m_s}{z^{2s+1}}\gk_{(\al_1\al_2}\dots\gk_{\al_{2s-1}\al_{2s)}}\,.
\ee
Therefore, $f$ is no more than polynomial for a finite set of
spins, despite the fact that the twisted-adjoint module $\C$ is
infinite dimensional. Note that \eqref{Dtype} are of Petrov $D$
type and as pointed out in \cite{Didenko:2021vui} have multi-copy
origin.

The scalar case $s=0$ corresponds to
\be\label{D1}
C^{(1)}_{\Delta=1}=m_0 z\,e^{i y_{\al}\bar y^{\al}}\,,
\ee
which is the solution with $\Delta=1$ boundary $z=0$ asymptotics.
Note, that \eqref{Csol1} contains no scalar with alternative
boundary condition $\Delta=2$. It was also found in
\cite{Didenko:2021vui} in the form
\be\label{D2}
C^{(1)}_{\Delta=2}=m_0'{z^2}(1+i y_{\al}\bar y^{\al})e^{i
y_{\al}\bar y^{\al}}\,.
\ee
We exclude $\Delta=2$ contribution from consideration by setting
$m_0'=0$ for simplicity. It is important to stress that
\eqref{Csol1}, \eqref{wsol} is a solution of the {\it type--A} HS
model \eqref{A} in \eqref{OMS} and at higher orders \eqref{CCeq}
and is not applicable for general $\eta$.

In terms of current module \eqref{CT1}, solution \eqref{Csol1} and
\eqref{wsol} corresponds to the following (anti)holomorphic
currents
\begin{align}
&\T^{b}_{(1)}=(f(w)+f(\bar w))(k+\bar k)\,,\label{bosT}\\
&\T^{h}_{(1)}=f(w)k+f(\bar w)\bar k\,,\label{chT}
\end{align}
where
\be\label{f}
f(w):=f\left(\ff{z}{2}\gk_{\al\gb}w^{\al}w^{\gb}\right)\,,\qquad
f(\bar w):=f\left(\ff{z}{2}\gk_{\al\gb}\bar w^{\al}\bar
w^{\gb}\right)
\ee
and $\T^{b}_{(1)}$ gives the standard bosonic embedding
\eqref{bosembd}, while $\T^{h}_{(1)}$ the chiral one \eqref{Cwch}.
Note that \eqref{bosT} and \eqref{chT} both being the sum of
holomorphic and antiholomorphic functions trivially solve
\eqref{Teq}. Indeed, since $\nabla f(w)=\nabla f(\bar w)=0$,
\eqref{Teq} is satisfied because it contains mixed derivatives.
Gauge 1-form $\w$ is expressed via \eqref{wsol} for both
embeddings as follows
\be
\w_{(1)}^{b, h}=-\ff{i}{2}{\bf
E}^{\al\gb}\int_{0}^{1}\dr\tau\p_{\al}\p_{\gb}f(\tau
w+(1-\tau)\bar w)(1+k\bar k)\label{wsol1}
\ee
with $f(w)$ given by \eqref{f}. Note, that in terms of currents
solution \eqref{Csol1} looks very simple being just power $s$ of
covariantly constant $\ff{z}{2}\gk_{\al\gb}w^{\al}w^{\gb}$ for
spin $s$.

\subsection{Quadratic analysis} Second order corrections of the
twisted-adjoint sector are governed by \eqref{TT}, where its right
hand side should be computed on free solutions \eqref{bosT} and
\eqref{wsol1} in the bosonic case and \eqref{chT}, \eqref{wsol1}
in the case of chiral embedding. This implies that quadratic
source $\J^c(\T^{(1)},\T^{(1)})+\J^{\go}(\T^{(1)}, \w^{(1)})$ is
totally defined in terms of function $f(x)$. Let us calculate it
in the bosonic case \eqref{bosT}. For that we take
\be\label{T2}
\T^{b}_{(2)}=T(w, \bar w; z)(k+\bar k)\,.
\ee
Substituting \eqref{T2} into the left hand side of \eqref{TT}
while \eqref{bosT} and \eqref{wsol1} into the right one and after
straightforward calculation we arrive at
\be\label{maineq}
\nabla T+i\, {\bf E}^{\al\gb}\,\p_{\al}\bar\p_{\gb}
T+\ff{i}{2}{\bf E}\,\p_{\al}\bar\p^{\al}T={\bf
E}^{\al\gb}J_{\al\gb}+{\bf E}\,J\,,
\ee
where
\be\label{Jdecomp}
J_{\al\gb}=J^c_{\al\gb}+J^{\go}_{\al\gb}\,.
\ee
$J^c_{\al\gb}$ and $J$ come from current terms \eqref{JCCaa} and
\eqref{JCC} correspondingly
\begin{align}\label{Jffaa}
&J^{c}_{\al\gb}=\frac{1}{2z^2}\int \ff{\dr u\dr
v}{(2\pi)^2}\int_{0}^{1}\dr\tau e^{\ff{i}{z}u_{\al}
v^{\al}}\Big[-w_{\al}\ff{\p}{\p w^{\gb}}(f(v+\bar
w+(1-\tau)w)f(u+\bar w-\tau w))+\nn\\
&+\bar w_{\al}\ff{\p}{\p\bar w^{\gb}}(f(u+y+(1-\tau)\bar
w)f(v+w-\tau\bar w))\Big]\\
&+\frac{1}{2}\int_{0}^{1}\dr\tau\Big[-w_{\al}f(\tau y)\ff{\p}{\p
w^{\gb}}(f(\bar w+\tau w)+f(\bar w-\tau w))+\bar w_{\al}f(\tau
\bar w)\ff{\p}{\p \bar w^{\gb}}(f(w+\tau\bar w)+f(w-\tau\bar
w))\Big]+\nn\\
&+(\al\leftrightarrow\gb)\nn
\end{align}
and
\begin{align}\label{Jff}
&J=\frac{1}{2z^2}\int \ff{\dr u\dr v}{(2\pi)^2}\int_{0}^{1}\dr\tau
e^{\ff{i}{z}u_{\al} v^{\al}}\Big[w^{\al}\ff{\p}{\p
w^{\al}}(f(v+\bar
w+(1-\tau)w)f(u+\bar w-\tau w))+\\
&+\bar w^{\al}\ff{\p}{\p\bar w^{\al}}(f(u+w+(1-\tau)\bar
w)f(v+w-\tau\bar w))+ z^2 f(\tau w)w^{\al}\ff{\p}{\p w^{\al}} (f(\bar
w+\tau w)+f(\bar w-\tau w))+\nn\\
&+ z^2 f(\tau\bar w)\bar w^{\al}\ff{\p}{\p\bar w^{\al}} (f(w+\tau\bar
w)+f(w-\tau\bar w))\Big]\,,\nn
\end{align}
while $J^\go$ comes from \eqref{JwC}
\begin{align}\label{Jgob}
&J^{\go}_{\al\gb}=\frac{i}{z}\int \ff{\dr u\dr
v}{(2\pi)^2}\int_{0}^{1}\dr\tau e^{\ff{i}{z}u_{\al}
v^{\al}}\ff{\p}{\p\bar w^{\al}}\ff{\p}{\p\bar
w^{\beta}}\Big(f(v+w)(f(w+\bar w+\tau u)-f(\bar w-w+\tau u))\Big)+\nn\\
&+e^{-\ff{i}{z}\bar u_{\al}\bar v^{\al}}\ff{\p}{\p
w^{\al}}\ff{\p}{\p w^{\beta}}\Big(f(\bar v+\bar w)(f(w+\bar w+\tau
\bar u)-f(w-\bar w+\tau\bar u))\Big)\,.
\end{align}
In obtaining the above expressions we use that $f(y)$ is even and
also change integration variable $\tau\to 1-\tau$ in certain
terms. Compatibility condition $\nabla^2=0$ for \eqref{maineq}
gives the following constraint on the interaction terms
\be\label{consist}
\frac{i}{2}({\bf E}^{\al\gb}\p_{\al}\bar\p_{\gb}+{\bf
E}\p_{\al}\bar\p^{\al})({\bf E}^{\gga\gd}J_{\gga\gd}+{\bf
E}J)=-{\bf E}^{\al\gb}\nabla J_{\al\gb}\,,
\ee
where in obtaining this result we applied $\nabla$ to
\eqref{maineq} and noted that
\be\label{pr11}
{\bf E}\,\nabla J=0\,.
\ee
The latter property is due to the fact that since $\nabla f=0$,
$\nabla$ acts on \eqref{Jff} nontrivially in its manifest $z$ --
dependence only. The result of that action is therefore
proportional to $\nabla z={\bf E}$ and since ${\bf E}^2=0$,
\eqref{pr11} holds. Using the Schouten identity
\be
{\bf E}^{\al\gb}{\bf E}^{\gga\gd}=\half \mathbf{E}^\alpha {}_\mu \mathbf{E}^{\mu\delta} \epsilon^{\beta \gamma}+\half \mathbf{E}^\beta {}_\mu \mathbf{E}^{\mu\gamma}\epsilon^{\alpha \delta}
\ee
one rewrites \eqref{consist} further as
\begin{align}
&\p_{\gga}\bar\p_{\al}J^{\gga}{}_{\gb}+\bar\p_{\gga}\p_{\al}J^{\gga}{}_{\gb}=0\,,\label{pr1}\\
&\ff i2 {\bf
E}(\p_{\al}\bar\p_{\al}J-\p_{\gb}\bar\p^{\gb}J_{\al\al})=\nabla
J_{\al\al}\,.\label{pr2}
\end{align}
\eqref{pr1} and \eqref{pr2} are satisfied as a result of the
consistency of HS equation \eqref{CCeq} and needs no verification.

\subsubsection{Static condition}

Let us now note, that since we are interested in the deformation
of the linearized solution \eqref{bosT} by HS interactions, the
former should depend on the same set of fields as the latter. This
implies that the corresponding $T$ in \eqref{T2} being a solution
of \eqref{maineq} should depend on the Lorentz components of the
global symmetry parameter $K_{AB}$, \eqref{param} only, much as
the free solution does. Those are the $\nabla$-- covariantly
constant upon proper rescaling fields \eqref{const0} and Lorentz
scalar $z$. Thus, the only allowed space-time dependence for $T$
which is not constant with respect to $\nabla$ is via variable
$z$. This guarantees that the deformed solution maintains planar
and static symmetry. In particular, there must be no dependence on
time $t$ \eqref{time} in $T$ either which is the static
requirement. Practically this means that $\nabla T\sim {\bf E}$
contains no one-forms ${\bf E}_{\al\gb}$. This in turn implies
that \eqref{maineq} can be equivalently rewritten as
\begin{align}
&\ff12\left(\p_{\al}\bar\p_{\gb}+\p_{\gb}\bar\p_{\al}\right)T=-i\,J_{\al\gb}\,,\label{plain}\\
&\left({\nabla}+\ff{i}{2}{\bf
E}\,\p_{\al}\bar\p^{\al}\right)T={\bf E}\,J\,.\label{radial}
\end{align}
Eq. \eqref{plain} governs the dependence of $T$ on $\nabla$ --
constant fields, while \eqref{radial} its radial
dependence\footnote{To be a bit more specific, using the
Poincar\'{e} coordinates \eqref{Poincare} one shows that
\eqref{plain} defines behavior on boundary coordinates $x^i$ in a
way consistent with planar symmetry, while \eqref{radial} sets
evolution inside the bulk along $z$.} on $z$. Note that
\eqref{plain} places stringent constraints on possible HS
interactions that respect nonlinear deformation \eqref{T2} and
will be called the {\it static condition}. Its relation to static
symmetry is as follows. Suppose $T$ depends also on time, i.e.,
$T=T(w, \bar w; z, t)$. In that case property $\nabla T\sim{\bf
E}$ is no longer true, since $\nabla T=\ff{\p}{\p z}T\,{\bf
E}+\ff{\p}{\p t}T\,\dr t$, where $\dr t$ is given by \eqref{time},
which contains ${\bf E}^{\al\gb}$ -- part. So, one can not
separate \eqref{maineq} into pieces \eqref{plain} and
\eqref{radial}.

HS interactions that are consistent with \eqref{plain} should
satisfy the following compatibility conditions
\begin{align}
&\p_{\gga}\bar\p_{\al}J^{\gga}{}_{\gb}+\bar\p_{\gga}\p_{\al}J^{\gga}{}_{\gb}=0\,,\label{c1}\\
&\p_{\al}\p_{\gb}J^{\al\gb}=\bar\p_{\al}\bar\p_{\gb}J^{\al\gb}=0\,.\label{c2}
\end{align}
While \eqref{c1} is already on the list \eqref{pr1} being a part
of general HS consistency, it is not granted that \eqref{c2} is
satisfied and therefore should be verified separately.
Verification of \eqref{c2} being quite tedious is the most
challenging part of this paper. The important result that ensures
the existence of planar solution is that \eqref{c2} does hold for
bosonic embedding \eqref{T2}. For chiral embedding \eqref{chT}
static constraint \eqref{c2} turns out to be {\it not} satisfied.
We interpret this result as there is no quadratic deformation of
the linearized solution \eqref{chT} which is both static and
planar in the chiral case. Though rather straightforward, the
check of \eqref{c2} involves few partial integrations and we leave
it for appendix.

\subsection{Solving the bosonic embedding}

Once \eqref{c2} is shown to be satisfied for bosonic embedding
\eqref{T2}, we can try to solve \eqref{plain} and \eqref{radial}
and find this way the nonlinear deformation of planar solution
\eqref{bosT} in the sector of HS curvatures. To do so we note that
\eqref{plain} can be rewritten as
\be\label{ddeq}
\p_{\al}\bar\p_{\gb}T=-i\,J_{\al\gb}+i\gep_{\al\gb}
h:=X_{\al\gb}\,,
\ee
where $h=h(w, \bar w; z)$ is some unknown function. Indeed,
\eqref{plain} says that the symmetrized derivatives acting on $T$
gives $J_{\al\gb}$. This statement is equivalent to the following.
Relaxing the symmetrization of the derivatives one still
reproduces $J_{\al\gb}$ up to something which is skew symmetric
$h_{\al\gb}=-h_{\gb\al}$. But since spinorial indices range only
two values, $h_{\al\gb}\sim\gep_{\al\gb}$ and therefore we have
\eqref{ddeq}. Function $h$ can not be arbitrary however as it is
constrained by the following integrability conditions
\be\label{heq}
\p_{\al}h=-\p_{\gb}J_{\al}{}^{\gb}\,,\qquad \bar\p_{\al}
h=\bar\p_{\gb}J_{\al}{}^{\gb}\,.
\ee
Consistency of \eqref{heq} in turn is fulfilled by \eqref{c2}. The
general solution of \eqref{ddeq} can be written down by applying
two consecutive contracting homotopies
\be\label{Tsol}
T=\int_{[0,1]^2}\dr\tau\dr\bar\tau
w^{\al}\bar{w}^{\gb}X_{\al\gb}(\tau w, \bar\tau\bar w;
z)+\phi(w;z)+\bar\phi(\bar w;z)\,,
\ee
where $\phi(w;z)$ and $\bar \phi(\bar w; z)$ are arbitrary at this
stage (anti)holomorphic functions. The $\phi$--part of $T$ in
\eqref{Tsol} encodes HS Weyl tensors that are extracted from $T(w,
0; z)$ and $T(0, \bar w; z)$, while the $X$--part corresponds to
their on-shell derivatives. Hence, physical information about the
nonlinear HS deformation is stored in yet unknown
$\phi$--functions. Note that the scalar contribution to $T$ is
placed within sum $\phi(0,z)+\bar\phi(0,z)$ rather than in the
individual pieces $\phi(0,z)$ and $\bar\phi(0,z)$. This allows one
to set
\be\label{phicond}
\phi(0,z)=\bar\phi(0,z)
\ee
for convenience. The solution of \eqref{ddeq} can be explicitly
found once $h(w,\bar w; z)$ is known. To restrict the latter one
can again use contracting homotopy in both equations \eqref{heq}.
This will fix $h(w,\bar w; z)$ up to a $w$ and $\bar
w$--independent function of $z$. Indeed, from first equation in
\eqref{heq} one finds
\be\label{h1}
h=-\int_{0}^{1}\ff{\dr\tau}{\tau}w^{\al}\ff{\p}{\p
w^{\gb}}J_{\al}{}^{\gb}(\tau w,\bar w)+\bar g(\bar w; z)\,,
\ee
while from the second,
\be\label{h2}
h=\int_{0}^{1}\ff{\dr\tau}{\tau}\bar w^{\al}\ff{\p}{\p \bar
w^{\gb}}J_{\al}{}^{\gb}(w,\tau\bar w)+g(w; z)\,,
\ee
where $g(w; z)$ and $\bar g(\bar w; z)$ are so far unspecified
functions. Since \eqref{h1} and \eqref{h2} should give one and the
same result for $h(w,\bar w; z)$ by equating the two expressions
and setting either $w$ or $\bar w$ to zero we fix $g$ and $\bar g$
up to an arbitrary constant ${\bf c}(z)$ implying the following
final result
\be\label{hf1}
h=\int_{0}^{1}\ff{\dr\tau}{\tau}\left(\bar w^{\al}\ff{\p}{\p \bar
w^{\gb}}J_{\al}{}^{\gb}(w,\tau\bar w; z)-w^{\al}\ff{\p}{\p
w^{\gb}}J_{\al}{}^{\gb}(\tau w,0; z)\right)+{\bf c}(z)
\ee
or equivalently
\be\label{hf2}
h=\int_{0}^{1}\ff{\dr\tau}{\tau}\left(-w^{\al}\ff{\p}{\p
w^{\gb}}J_{\al}{}^{\gb}(\tau w,\bar w; z)+\bar w^{\al}\ff{\p}{\p
\bar w^{\gb}}J_{\al}{}^{\gb}(0,\tau\bar w; z)\right)+{\bf c}(z)\,.
\ee
That \eqref{hf1} and \eqref{hf2} are identical is guaranteed by
consistency condition \eqref{c2}. The result of the analysis of
\eqref{plain} allows one fixing $T$ up to an arbitrary holomorphic
function $\phi(w; z)$ (and its complex conjugate) and $w, \bar w$
-- independent function ${\bf{c}}(z)$. Explicit expression for
$h(w,\bar{w})$ is provided in the appendix.

\subsubsection{Determining ${\bf c}(z)$} It is not difficult to
obtain equation on ${\bf c}(z)=h(0,0; z)$. To do so we first note
that from \eqref{ddeq} it follows
\be\label{ddT}
\p_{\al}\bar\p^{\al}T=2ih\,.
\ee
Acting then on \eqref{ddeq} with $\nabla$ and using \eqref{radial}
one gets
\be
{\bf
E}\,\p_{\al}\bar\p_{\gb}(J-\ff{i}{2}\p_{\gga}\bar\p^{\gga}T)=i\nabla(-J_{\al\gb}+\gep_{\al\gb}h)\,,
\ee
which upon substitution \eqref{ddT} gives
\be\label{Jheq}
{\bf E}\,\p_{\al}\bar\p_{\gb}(J+
h)=i\nabla(-J_{\al\gb}+\gep_{\al\gb}h)\,.
\ee
Contracting indices in \eqref{Jheq} and using that
$J_{\al\gb}=J_{\gb\al}$ we then have
\be\label{Jheqtr}
{\bf E}\,\p_{\al}\bar\p^{\al}(J+h)=2i\nabla h\,.
\ee
From  \eqref{heq} one extracts
\be
\p_{\al}\bar\p^{\al} h=-\p_{\al}\bar\p_{\gb}J^{\al\gb}\,,
\ee
which yields
\be
\nabla h=\ff{1}{2i}{\bf E}\,(\p_{\al}\bar\p^{\al}
J-\p_{\al}\bar\p_{\gb}J^{\al\gb})\,.
\ee
Finally, setting $w_{\al}=\bar w_{\al}=0$ and using \eqref{hf1} we
arrive at the desired equation for ${\bf{c}}(z)$
\be\label{ceq}
{\bf c}'(z)=\ff{1}{2i}(\p_{\al}\bar\p^{\al} J(0, 0;
z)-\p_{\al}\bar\p_{\gb}J^{\al\gb}(0, 0; z))\,.
\ee
The right hand side is readily accessible from
\eqref{Jffaa}-\eqref{Jgob} and can be easily calculated
explicitly. Quite interestingly, the result appears to have a neat
form of the total derivative with respect to $z$ once one takes
into account the specific dependence $f(w)$ on $w$ via a
combination \eqref{f}
\be
x=\ff{z}{2}\gk_{\al\gb}w^{\al}w^{\gb}\,.
\ee
One then finds
\be
{\bf c}'(z)=2\ff{\p}{\p z} \int \ff{\dr u\dr
v}{(2\pi)^2}\, e^{\ff{i}{z}u_{\al}v^{\al}}f'(v)f'(u)\,,
\ee
where $f'(w):=\ff{\p}{\p x} f(w)$. Solution for ${\bf c}(z)$ has a
freedom from homogeneous equation that gives some constant ${\bf
c}_0$. It can be set to zero, because ${\bf c}_0\neq 0$
corresponds to a freedom in solution of homogeneous field equation
\eqref{Teq}, which has been fixed at free level already. Indeed,
as is seen from \eqref{Tsol}, nonzero ${\bf c}_0$ induces free
scalar with $\Delta=2$ boundary behavior \eqref{D2}
\be
T_{\Delta=2}=\,\,{\bf c}_0(z+iw_{\al}\bar w^{\al})\,.
\ee
This has been set to zero at free level. As a result,
\be\label{cz}
{\bf c}(z)=2\int \ff{\dr u\dr v}{(2\pi)^2}\,
e^{\ff{i}{z}u_{\al}v^{\al}}f'(v)f'(u)\,.
\ee

\subsubsection{HS Weyl tensors}

Having found ${\bf c}(z)$ we have completely determined function
$h$ that enters \eqref{ddeq} via \eqref{hf1}. Using this
information we can now find the leftover $\phi(w; z)$ and $\bar
\phi(\bar w; z)$ from \eqref{Tsol} thus fully fixing the current
module at second order and correspondingly the twisted-adjoint one
through \eqref{CT1}. Recall, that all HS Weyl tensors are stored
in $\phi$'s. To arrive at the equation on $\phi$ we substitute
\eqref{ddT} into \eqref{radial} to obtain
\be\label{nablaT}
\nabla T={\bf E}\,(J+h)\,.
\ee
Feeding \eqref{Tsol} and \eqref{hf1}, \eqref{cz} into
\eqref{nablaT} one gets pretty complicated equation, which
contains all the required information about $\phi$'s. Equation
that determines evolution of $\phi(w; z)$ along $z$ can be easily
extracted from it by setting $\bar w_{\al}=0$
\be\label{Weq1}
{\p_z} (\phi(w; z)+\bar \phi(0; z))=J(w, 0; z)+ h(w, 0; z)\,.
\ee
Using now \eqref{phicond} and also that $J(0,0; z)=0$ one finds
\be
2\phi'(0; z)=h(0, 0; z)={\bf c}(z)\,,
\ee
hence,
\be\label{Weq}
\p_z \phi(w; z)=J(w, 0; z)+ h(w, 0; z)-\ff12{\bf c}(z)\,.
\ee
Similar equation arises for $\bar \phi(\bar w; z)$. Upon
straightforward if perhaps somewhat annoying calculation one is
able to arrive at quite a simple explicit form for the right hand
side of \eqref{Weq}
\begin{align}\label{Jh}
&J(w,0)+h(w,0)-\ff12{\bf c}(z)=\\
&=\frac{1}{z^2}\int \ff{\dr u\dr v}{(2\pi)^2}\int_{0}^{1}\dr\tau
e^{\ff{i}{z}u_{\al}v^{\al}}\Big(f(w+u)f(w+v)+f(v)f(u-w)-2f(u)f(v)-\nn\\
&-f(u-\tau w)\p_{\tau}f(v+(1-\tau)w)+z^2\,f'(u)f'(v)\Big)\,.\nn
\end{align}
Closed form solution of \eqref{Weq} can be written down as a
homotopy integral
\begin{align}
&\phi(w, z)=\int \ff{\dr u\dr
v}{(2\pi)^2}\int_{[0,1]^2}\dr\tau\dr\gs
e^{\ff{i}{z\gs}u_{\al}v^{\al}}\frac{1}{\sigma^2 z}\Big(f(w+u)f(w+v)+f(v)f(u-w)-2f(u)f(v)-\nn\\
&-f(u-\tau w)\p_{\tau}f(v+(1-\tau)w)+\sigma^2 z^2 \,f'(u)f'(v)\Big)\,.\nn
\end{align}
Poles $\ff{1}{\gs}$ in the exponential and in the prefactor may
look worrisome, but in fact the integral is perfectly well defined
as can be seen upon changing variables $u\to \sqrt{\sigma z } u$,
$v\to \sqrt{\gs z }v$
\begin{align}\label{phisol}
&\phi(w, z)=z\int \ff{\dr u\dr
v}{(2\pi)^2}\int_{[0,1]^2}\dr\tau\dr\gs
e^{{i}u_{\al}v^{\al}}\\
&\Big(f(w+u\sqrt{\gs z})f(w+v\sqrt{\gs z})+f(v\sqrt{\gs z })f(u\sqrt{\gs z }-w)-
2f(u\sqrt{\gs z })f(v\sqrt{\gs z })-\nn\\
&-f(\sqrt{\gs z }u-\tau w)\p_{\tau}f(\sqrt{\gs z }v+(1-\tau)w)+\sigma^2 z^2 \, f'(\sqrt{\gs z }u)f'(\sqrt{\gs z }v)\Big)\,.\nn
\end{align}
There is a natural freedom in $z$ -- independent solutions
$\phi_0(w)$ of homogeneous part of equation \eqref{Weq}. We set
$\phi_0(w)=0$ for it corresponds to the solutions of the free HS
equations which have been already fixed by \eqref{bosT}.

The full result for $T$, \eqref{Tsol} has a complicated form and
is not particularly illuminating to be written down here. Its most
important part, however, is stored in \eqref{phisol}. Indeed,
$\phi(w; z)$ is nothing but the generating function for quadratic
corrections to holomorphic components of the HS Weyl tensors as
follows from \eqref{CT1} for $s>0$ ($y\neq 0$)
\be\label{Weyl2}
C^{(2)}(y, 0; x)=z\phi(\sqrt{z}y; z)
\ee
and
\be
C^{(2)}(0, 0; x)=z(\phi(0; z)+\bar\phi(0;z))=2z\,\phi(0;z)\,.
\ee
for $s=0$.

\subsection{Properties of the quadratic corrections}\label{prop} Here we
collect some important properties of our solution.
\begin{itemize}
\item{{\bf Algebraically special}} Recall that planar solution
\eqref{feq}  is algebraically special for every spin $s>0$,
\eqref{feqs} at the linearized level. This is seen from
\eqref{Dtype}, which says that the HS Weyl tensors are of the
generalized Petrov $D$ -- type. The same is true for quadratic
deformation. Indeed, by analyzing \eqref{phisol} for $f=\sum_s
f_s$, \eqref{feqs} one can schematically extract
\be\label{Cres}
C_{\al_1\dots\al_{2s}}^{(2)}=\sum_{i,j} a(m_i, m_j; z)
\gk_{(\al_1\al_2}\dots\gk_{\al_{2s-1}\al_{2s})}\,,
\ee
where $a(m_i, m_j; z)$ are some bilinear $z$--dependent
coefficients of HS parameters $m_s$, \eqref{feqs}. Thus, the
algebraic type is again Petrov $D$.

\item{{\bf Spin finiteness}} Suppose $f(y)$ is polynomial
corresponding to a finite amount of nonzero HS spins at free
level, then the quadratic correction \eqref{phisol} contains no
more than finite number of planar HS excitation too. For example,
should one starts out with just an $s=2$ black brane at free level
\be\label{s2}
f_2(w)=\ff{m_2}{2}
\left(\ff{z}{2}\gk_{\al\gb}w^{\al}w^{\gb}\right)^2\,,
\ee
then the quadratic correction calculated from \eqref{phisol} cuts
short at $s=4$ given by the last term below
\be\label{phi2}
\phi(w; z)=\frac{1}{5} m_2^2 z^5+\frac{7}{6}m_2^2 z^3
\left(\ff{z}{2}\gk_{\al\gb}w^{\al}w^{\gb}\right)^2+\frac{71}{280}m_2^2 z
\left(\ff{z}{2}\gk_{\al\gb}w^{\al}w^{\gb}\right)^4\,.
\ee
This is the general feature of our solution. If $f(w)$ is bounded
by some spin $s$, then the deformation terminates at $s_{max}=2s$.
At the same time any such spin $s>0$ induces a tower of nonzero
spins $0,\dots, 2s$.

Note that while $s=2$ Weyl tensor \eqref{s2} is exactly the same
at the linearized level as of a GR black brane, this is no longer
so in HS interactions. From \eqref{phi2} it follows that not only
spin $s=2$ receives correction that brings it away from ordinary
black brane, but also the new fields $s=0$ and $s=4$ show up.

The corrections from the $s=2$ sector to itself are nonzero since
the gravity interaction differs from that of Einstein within the
HS theory. This leads to the breakdown of the planar GR black hole
\eqref{s2} already by quadratic HS corrections.

\item{{\bf Multi-copy structure}}

Since by construction solution \eqref{phisol} is a deformation of
the linearized solution, which has a structure of the double copy,
it enjoys the double copy or better put it the multi-copy property
too. Indeed, as is seen from \eqref{Cres}, each spin $s$ HS Weyl
tensor is expressed via a multiple product of spin $s=1$ solution
\be\label{s1}
C_{\al\gb}=\ff12 z^3\gk_{\al\gb}\,,\qquad \bar C_{\dal\dgb}=\ff12
z^3\bar\gk_{\dal\dgb}\,,
\ee
where from \eqref{brane} $z$ itself is the zeroth copy of Maxwell
field
\be\label{s0}
z=(-2C_{\al\gb}C^{\al\gb})^{\ff14}\,.
\ee
Therefore, the spin $s$ HS Weyl tensor from \eqref{Cres} acquires
a form of a power $s$ of Maxwell field \eqref{s1} up to a $z$ --
dependent prefactor which can be expressed via zeroth copy
\eqref{s0}. The multi-copy results from Petrov $D$-type form of
the solution and its sufficiently reach global symmetry.

\item{{\bf Closed scalar sector}} While it follows from
\eqref{phisol} that every spin contribution in $f(y)$ induces
lower and higher spins as is generally expected for HS algebraic
reason, this is not so for $f=1$. This case corresponds to a
single $\Delta=1$ scalar $C=z$. Quadratic corrections to the spin
$s=0$ sector should vanish in general because scalar vertex
$0-0-0$ is absent in four dimensions \cite{Sezgin:2003pt} being
conformally invariant \cite{Gelfond:2015poa}. This is indeed what
happens for $f=1$ as one finds from \eqref{phisol} $\phi(0; z)=0$.
But it turns out that the result is even stronger than that. In
this case we have
\be
\phi(w; z)=0\,,
\ee
which means that there are no corrections from $s=0$ to spins
$s>0$ either. In other words, $f=1$ is an exact solution at this
order. It will be interesting to see whether it remains so at
higher orders.

\item{{\bf Higher spin scalar condensate}} An interesting property
of \eqref{phisol} comes from inspection of its scalar sector. Let
us take a general
\be
f(w)=\sum_{s=0}\ff{m_s}{s!}\left(\ff{z}{2}\gk_{\al\gb}w^{\al}w^{\gb}\right)^s
\ee
with {\it{a priori}} arbitrary HS real parameters $m_s$ and let us
find its contribution to the scalar sector from \eqref{phisol}.
Straightforward calculation gives
\be\label{pot}
\phi(0; z)=\sum_{s=1}\ff{m_s^2}{2s+1}z^{2s+1}\,.
\ee
Remarkably, the obtained contribution turns out to be diagonal in
spins and is strictly positive for $z>0$ no matter what signs of
HS charges $m_s$ are. So we see that HS fields at this order
condensate into a positive scalar. Note also that such a
correction affects neither free $\Delta=1$, nor $\Delta=2$ scalar
boundary behavior. It would be very interesting to understand if
there are any fundamental grounds that lead to positivity of
scalar perturbation.

An example of what \eqref{pot} could be if, lets say, all HS
parameters for every integer spin are equal to each other $m_s=m$
is
\be
\phi(0; z)= m^2 (\arctanh{z}-z)
\ee
leading to the following second order corrections to the scalar
field
\be
C^{(2)}(0,0; z)=2m^2 z\, (\arctanh{z}-z)=\ff{2}{3} m^2
z^4+O(z^6)\,.
\ee
\end{itemize}

\section{Concluding remarks}
In this paper we have completed our analysis of the bosonic planar
HS solutions at leading interaction order. At the linearized level
such solutions have been proposed in \cite{Didenko:2021vui} while
the nonlinear deformation in the sector of HS curvatures is found
here.

Solving nonlinear HS equations order by order is generally a
highly nontrivial task. The problem is twofold. First off, HS
equations contain higher derivatives which grow with spin making
HS equations quite complicated already at leading interaction
order. Second, it is not infrequently that even the simplest free
solutions have increasingly complicated with spin form, especially
those in the sector of gauge fields (see e.g.,
\cite{Bolotin:1999fa}, \cite{Nagaraj:2019zmk} for the plane wave
like examples in $AdS$). Both problems together normally square
difficulties of perturbative expansion.

What makes life easier in our case is a remarkably simple
linearized version of HS planar solution \eqref{Csol1},
\eqref{wsol}. It exists in the {\it type-A} HS theory only with
the phase parameter $\eta=1$ and we expect that a similar solution
away from this point would be much more complicated. Moreover,
being static and having planar symmetry the solution appears to be
perfectly suited for what we call current ansatz \eqref{CT1},
which allows one isolating HS currents from the twisted-adjoint
module. In terms of 'currents' each spin planar contribution
corresponds to a simple monomial. Originally introduced in
\cite{Vasiliev:2012vf} for HS bulk-boundary analysis at free
level, the current ansatz unexpectedly goes through in
interactions \cite{Didenko:2017lsn}. Having generalized this
ansatz to make it covariant with respect to planar symmetry, we
use it here at the nonlinear level and this has brought us extra
simplification.

A great deal of technical analysis is concerned with checking
whether the linearized solution admits nonlinear deformation or
not. Putting it differently, the question is whether the nonlinear
solution depends on the same set of fields as the free one.
Practically, this amounts to verifying static constraint
\eqref{c2} for particular HS interactions. We found that it is
satisfied for the standard bosonic HS model, while this may not be
the case for a different type of embeddings into the susy HS
system. Particular example of the so called chiral embedding
\eqref{Cwch} was shown not to respect the static constraint.

In the bosonic case we found quadratic corrections to HS Weyl
tensors in a closed form \eqref{Weyl2}. These turned out to be
algebraically special of Petrov type $D$. Moreover, in the spirit
of the Weyl double copy, the final result can be reconstructed
from the free spin $s=1$ solution polynomially using the zeroth
copy scalar for every spin $s$. Therefore the obtained solution
naturally generalizes the double copy form of GR black holes
observed in \cite{Didenko:2008va} (see also
\cite{Monteiro:2014cda}) to nonlinear HS interacting level.

The solution is characterized by an arbitrary real function $f(y)$
or equivalently by (in)finitely many real arbitrary HS parameters
$m_s$ that show up in its Taylor expansion and generalize the mass
parameter $m_2$ of a black brane. In particular, the case of a
finite number of $m_s$ corresponds to a polynomial $f(y)$. The
nonlinear corrections come out as bilinears in $m_s$. Their
properties are collected in section \ref{prop}.

Some notable features are as follows. While generally one may
expect that a given spin $s$ induces infinitely many spins in
interactions due to HS algebraic argument, in practice it did not
happen as quadratic corrections terminated at spin $2s$ for
kinematical reason. Particularly, spin $s=0$ yields no quadratic
corrections at all including to itself and thus forms a closed
interacting sector. Moreover, one may argue that there would be no
higher order corrections to $s>0$ from a scalar either due to
kinematical reasons as well.

Another interesting feature is HS contribution into scalar sector
\eqref{pot}. While as a matter of principle HS parameters can be
either positive or negative, their quadratic corrections to the
scalar turn out to be diagonal in $m_s$ and are strictly positive
manifesting universal behavior for all spins. It would be very
interesting to explore this phenomenon further.

Using \eqref{Csol1} as a free solution we left aside $\Delta=2$
scalar \eqref{D2}. We did that purposely as on one hand the extra
scalar brings in considerable complication of the analysis, yet
seemingly changing no qualitative picture of interactions. But on
the other, the whole scalar sector alone deserves a separate
consideration. The global symmetry is bigger for pure scalar than
the planar one considered in the presence of spinning fields. It
is given by the Poincar\'{e} symmetry of the three dimensional
boundary. Given that at the full nonlinear level a combination of
$\Delta=1$ and $\Delta=2$ scalars arguably induces no fields $s>0$
for kinematical reasons, it would be very interesting to find the
corresponding exact solution. If this is possible then one will be
able addressing questions on HS symmetry breaking by that
potentially simple yet very symmetric HS vacuum, which would be
just the $AdS_4$ in its gravity sector. It is intriguing to obtain
the mass spectrum that such a symmetry breaking may
deliver\footnote{We are grateful to Ruslan Metsaev for the very
useful discussion on this point.} (see \cite{Metsaev:1999kb} for a
proposal in this direction and also \cite{Barvinsky:2015wvz} for
phenomenological application of the conformal HS symmetry
breaking). We hope to consider this problem elsewhere.

Another comment concerns the double copy form of the solution.
Though the obtained leading order corrections appear to have the
multi-copy form, it was not granted in the first place. An example
of chiral embedding \eqref{Cwch} that has a multi-copy structure
at free level proves that it may not remain so in perturbations.
Indeed since the quadratic corrections fail to respect static
constaint \eqref{c2}, the solution depends on some extra fields
that are not present at first order. Therefore, the corresponding
HS Weyl tensors can not be expressed in terms of Maxwell tensor
\eqref{s1}.

Let us conclude by what our analysis adds to lower spin
interaction within HS theory. There is an interesting phenomenon
of GR black holes that make them linearly exact. For that reason
the $s=2$ component of the linearized solution \eqref{Csol1}
describes the standard black brane Weyl tensor. One could have
hoped that it may not be affected by HS interactions either, but
this is actually not happening. Not only $s=2$ gets corrected by
second order, it inevitably induces nontrivial $s=4$ field. More
generally, the lower spin system of $s=0,1,2$ that admits planar
symmetry is not closed under HS interactions for any values of
their charges.

\section*{Acknowledgments}
We are grateful to Ruslan Metsaev for useful discussion on
higher-spin symmetry breaking and to Mitya Ponomarev and M.A.
Vasiliev for valuable comments on the draft of the paper. We would
also like to thank Tim Adamo for correspondence. This research was
supported by the Russian Science Foundation grant 18-12-00507.

\section*{Appendix A. Verification of consistency condition}
In this section consistency condition \eqref{c2} is explicitly
checked for bosonic embedding \eqref{bosT}. Symmetric part of the
current can be treated as a sum of several contributions, namely
\begin{equation}\label{Jcomposition}
J_{\alpha \beta}={J}^{\go}_{\alpha \beta}+\widehat{J}^{\eta}
{}_{\alpha \beta}+\widehat{J}^{\bar{\eta}} {}_{\alpha \beta}
+\widetilde{J}^{\eta} {}_{\alpha \beta}+\widetilde{J}^{\bar{\eta}}
{}_{\alpha \beta}\,.
\end{equation}
We split $J^{c}_{\alpha \beta}$ into two terms compared to
\eqref{Jdecomp}. First term given by \eqref{Jgob} originates from
commutator of the first order correction to field $\mathbf{w}$
with field $\mathcal{C}$, while all other terms come from the
holomorphic and antiholomorphic vertices on $AdS$ background. Even
though our analysis requires to set phase parameter of the
Vasiliev theory to one \eqref{A}, it is useful to separate terms
proportional to $\eta$ from those proportional to $\bar\eta$ to
keep track of cancellations. Explicit expressions for all of the
contributions are provided below
\begin{multline}\label{first}
{J}^{\go}_{\alpha \beta}=\frac{i}{z} \int_0^1 dt\Bigg\{\int\frac{\dr u\, \dr v}{(2\pi)^2} e^{\frac{i}{z}u_\alpha v^\alpha} \, f(w+v) \frac{\partial^2}{\partial w^\alpha \partial w^\beta} f(y+\bar{y}+tu)+\\
+\int\frac{\dr \bar{u}\, \dr \bar{v}}{(2\pi)^2}
e^{-\frac{i}{z}\bar{u}_\alpha \bar{v}^\alpha} \,
f(\bar{y}+\bar{v}) \frac{\partial^2}{\partial w^\alpha \partial
w^\beta} f(y+\bar{y}+t\bar{u})-\\-
\int\frac{\dr u\, \dr v}{(2\pi)^2} e^{\frac{i}{z}u_\alpha v^\alpha}\, f(y+u)\frac{\partial^2}{\partial w^\alpha \partial w^\beta} f(y-\bar{y}+tv)-\\
-\int\frac{\dr \bar{u}\, \dr \bar{v}}{(2\pi)^2}
e^{-\frac{i}{z}\bar{u}_\alpha \bar{v}^\alpha}
f(\bar{y}+\bar{u})\frac{\partial^2}{\partial w^\alpha \partial
w^\beta} f(y-\bar{y}-t\bar{v})\Bigg\},
\end{multline}
\begin{multline}
\widehat{J}^{\eta} {}_{\alpha \beta}=-\frac{1}{2}\int_0^1
dt\Bigg\{w_\alpha f(tw) \frac{\partial}{\partial w^\beta} \big(
f(\bar{w}-tw)+f(\bar{w}+tw)\big)+w_\beta f(tw)
\frac{\partial}{\partial w^\alpha} \big(
f(\bar{w}-tw)+f(\bar{w}+tw)\big)\Bigg\},
\end{multline}
\begin{multline}
\widehat{J}^{\bar{\eta}} {}_{\alpha \beta}=\frac{1}{2}\int_0^1
dt\Bigg\{\bar{w}_\alpha f(t\bar{w}) \frac{\partial}{\partial
\bar{w}^\beta} \big(
f(w-t\bar{w})+f(w+t\bar{w})\big)+\bar{w}_\beta f(t\bar{w})
\frac{\partial}{\partial \bar{w}^\alpha} \big(
f(w-t\bar{w})+f(w+t\bar{w})\big)\Bigg\},
\end{multline}
\begin{multline}
\widetilde{J}^{\eta} {}_{\alpha
\beta}=-\frac{1}{2z^2}\left(w_\alpha \frac{\partial}{\partial
w^\beta}+w_\beta \frac{\partial}{\partial w^\alpha}\right)\int_0^1
dt \int\frac{\dr \bar{u} \, \dr \bar{v}}{(2\pi)^2}
e^{-\frac{i}{z}\bar{u}_\alpha\bar{v}^\alpha}f(\bar{w}+(1-t)w+\bar{u})
f(\bar{w}+\bar{v}-tw),
\end{multline}
\begin{multline}\label{last}
\widetilde{J}^{\bar{\eta}} {}_{\alpha
\beta}=\frac{1}{2z^2}\left(\bar{w}_\alpha
\frac{\partial}{\partial \bar{w}^\beta}+\bar{w}_\beta
\frac{\partial}{\partial \bar{w}^\alpha}\right)\int_0^1 dt \int
\frac{\dr u\, \dr v}{(2\pi)^2} e^{\frac{i}{z} u_\alpha
v^\alpha}f(w+(1-t)\bar{w}+u) f(w+v-t\bar{w}).
\end{multline}
To proceed with consistency check we apply derivatives to all of
the contributions and contract respective indices. Using partial
integration and Schouten identities, the results of
differentiation of each contribution \eqref{first}-\eqref{last}
are the following
\begin{multline}
\frac{\partial}{\partial w^\beta} J^{\go} {}_\alpha
{}^\beta(w,\bar{w})=-\frac{1}{z^2} \int \frac{\dr u \, \dr
v}{(2\pi)^2} e^{\frac{i}{z}u_\alpha v^\alpha}\,
f(w+v) \frac{\partial}{\partial w^\alpha} f(w+\bar{w}+u)-\\
-\frac{1}{z^2}\int \frac{\dr u\, \dr v}{(2\pi)^2}
e^{\frac{i}{z}u_\alpha v^\alpha}\, f(w+u) \frac{\partial}{\partial
w^\alpha} f(w-\bar{w}+v) +\frac{\partial}{\partial \bar{w}^\alpha}
\Big(f(w)\big(f(w+\bar{w})-f(w-\bar{w})\big)\Big),
\end{multline}
\begin{multline}
\frac{\partial}{\partial w^\beta} \widetilde{J}^{\bar{\eta}}
{}_\alpha {}^\beta(w,\bar{w})=\frac{1}{z^2}
\frac{\partial}{\partial \bar{w}^\alpha}\Bigg[ \int\frac{\dr u \,
\dr v}{(2\pi)^2} e^{\frac{i}{z}u_\alpha v^\alpha}\,
f(w+u+\bar{w}) f(y+v)-\\-\int\frac{\dr u \, \dr v}{(2\pi)^2} e^{\frac{i}{z}u_\alpha v^\alpha} \, f(w+u) f(w+v-\bar{w})\Bigg]-\\
-\frac{1}{z^2}\frac{\partial}{\partial w^\alpha} \int_0^1 dt \int \frac{\dr u \, \dr v}{(2\pi)^2} e^{\frac{i}{z}u_\alpha v^\alpha}\, f(w+u+(1-t)\bar{w}) f(w+v-t\bar{w})-\\
-\frac{1}{2z^2} \frac{\partial}{\partial
w^\alpha}\left(\bar{w}^\sigma \frac{\partial}{\partial
\bar{w}^\sigma}\right)\int_0^1 dt \int \frac{\dr u \, \dr
v}{(2\pi)^2} e^{\frac{i}{z}u_\alpha v^\alpha}\,
f(w+u+(1-t)\bar{w}) f(w+v-t\bar{w}),
\end{multline}
\begin{multline}
\frac{\partial}{\partial w^\beta} \widetilde{J}^{\eta} {}_\alpha {}^\beta(w,\bar{w})=-\frac{1}{z^2}\frac{\partial}{\partial w^\alpha} \int_0^1 dt \int\frac{\dr \bar{u} \, \dr \bar{v}}{(2\pi)^2} e^{-\frac{i}{z}\bar{u}_\alpha \bar{v}^\alpha}f(\bar{w}+(1-t)w+\bar{u})f(\bar{w}+\bar{v}-tw)-\\
-\frac{1}{2z^2} \frac{\partial}{\partial w^\alpha} \left(w^\sigma
\frac{\partial}{\partial w^\sigma}\right) \int_0^1 dt \int
\frac{\dr \bar{u}\, \dr \bar{v}}{(2\pi)^2}
e^{-\frac{i}{z}\bar{u}_\alpha\bar{v}^\alpha}\, f(\bar{w}+(1-t)
w+\bar{u}) f(\bar{w}+\bar{v}-tw).
\end{multline}
\begin{multline}
\frac{\partial}{\partial w^\beta} \widehat{J}^{\eta} {}_\alpha {}^\beta(w,\bar{w})=\frac{1 }{2}\frac{\partial}{\partial w^\alpha}  \int_0^1 dt\,  t  \left(w^\sigma \frac{\partial}{\partial \bar{w}^\sigma}\right) f(tw)\Big(f(\bar{w}+tw)-f(\bar{w}-tw)\Big)-\\
-{\eta }\frac{\partial}{\partial
\bar{w}^\alpha}f(w)\Big(f(\bar{w}+w)-f(\bar{w}-w)\Big),
\end{multline}
\begin{equation}
\frac{\partial}{\partial w^\beta} \widehat{J}^{\bar{\eta}}
{}_\alpha {}^\beta(w,\bar{w})=\frac{1}{2}\frac{\partial }{\partial
w^\alpha}\int_0^1 dt \, t  \left(\bar{w}^\sigma
\frac{\partial}{\partial w^\sigma}\right)
\Big[f(t\bar{w})\big(f(w+t\bar{w})-f(w-t\bar{w})\big)\Big].
\end{equation}
Summing up all the contributions one obtains
\begin{multline}\label{h1Exp}
\partial_{\beta} J_{\alpha} {}^\beta (w,\bar{w})=\frac{\partial}{\partial w^\alpha}\Bigg[-\frac{1}{z^2} \int_0^1 dt \int \frac{\dr u\, \dr v}{(2\pi)^2} e^{\frac{i}{z}u_\alpha v^\alpha}\, f(w+u+(1-t)\bar{w}) f(w+v-t\bar{w})-\\
-\frac{1}{z^2}\int_0^1 dt \int \frac{\dr \bar{u}\, \dr \bar{v}}{(2\pi)^2} e^{-\frac{i}{z}\bar{u}_\alpha \bar{v}^\alpha}f(\bar{w}+(1-t)w+\bar{u})f(\bar{w}+\bar{v}-tw)-\\
-\frac{1}{2z^2} \left(\bar{w}^\sigma \frac{\partial}{\partial \bar{w}^\sigma}\right)\int_0^1 dt \int\frac{\dr u\, \dr v}{(2\pi)^2} e^{\frac{i}{z}u_\alpha v^\alpha}\, f(w+u+(1-t)\bar{w}) f(w+v-t\bar{w})-\\
-\frac{1}{2z^2}  \left(w^\sigma \frac{\partial}{\partial w^\sigma}\right) \int_0^1 dt \int \frac{\dr \bar{u}\, \dr \bar{v}}{(2\pi)^2} e^{-\frac{i}{z}\bar{u}_\alpha\bar{v}^\alpha}\, f(\bar{y}+(1-t) y+\bar{u}) f(\bar{y}+\bar{v}-ty)+\\
+\frac{1}{2}  \int_0^1 dt\,  t  \left(w^\sigma \frac{\partial}{\partial \bar{w}^\sigma}\right) f(tw)\Big(f(\bar{w}+tw)-f(\bar{w}-tw)\Big)+\\
+\frac{1}{2}\int_0^1 dt \, t  \left(\bar{w}^\sigma
\frac{\partial}{\partial w^\sigma}\right)
\Big[f(t\bar{w})\big(f(w+t\bar{w})-f(w-t\bar{w})\big)\Big]\Bigg].
\end{multline}
Analogously one can compute contraction with barred derivative
\begin{multline}\label{h2Exp}
\bar{\partial}_\beta J_\al {}^\beta (w,\bar{w})=\frac{\partial}{\partial \bar{w}^\al}\Bigg[\frac{1}{z^2}  \int_0^1 dt \int\frac{\dr \bar{u}\, \dr \bar{v}}{(2\pi)^2} e^{-\frac{i}{z}\bar{u}_\alpha\bar{v}^\alpha} f(\bar{w}+(1-t)w+\bar{u})f(\bar{w}+v-tw)+\\
+\frac{1}{z^2}  \int_0^1 dt \int \frac{\dr u\, \dr v}{(2\pi)^2} e^{\frac{i}{z}u_\alpha v^\alpha}\, f(w+u+(1-t)\bar{w}) f(w+v-t\bar{w})+\\
+\frac{1}{2z^2} \left(w^\sigma \frac{\partial}{\partial w^\sigma}\right)\int_0^1 dt \int \frac{\dr \bar{u}\, \dr \bar{v}}{(2\pi)^2} e^{-\frac{i}{z}\bar{u}_\alpha\bar{v}^\alpha}\, f(\bar{w}+(1-t)w+\bar{u}) f(\bar{w}+\bar{v}-tw)+\\
+\frac{1}{2z^2} \left(\bar{w}^\sigma \frac{\partial}{\partial \bar{w}^\sigma}\right)\int_0^1 dt \int\frac{\dr u\, \dr v}{(2\pi)^2} e^{\frac{i}{z}u_\alpha v^\alpha}\, f(w+u+(1-t)\bar{w}) f(w+v-t\bar{w})-\\
-\frac{1}{2} \int_0^1 dt \, t  \left(w^\sigma \frac{\partial}{\partial \bar{w}^\sigma}\right)\Big[f(tw)\big(f(\bar{w}+tw)-f(\bar{w}-tw)\big)\Big]-\\
-\frac{1}{2} \int_0^1 dt \, t  \left(\bar{w}^\sigma
\frac{\partial}{\partial w^\sigma}\right)
\Big[f(t\bar{w})\big(f(w+t\bar{w})-f(w-t\bar{w})\big)\Big]\Bigg].
\end{multline}
After bringing the right hand sides of \eqref{h1Exp} and
\eqref{h2Exp} to the form of total derivatives, it is clear that
consistency conditions \eqref{c1} and \eqref{c2} are both
trivially satisfied.

\section*{Appendix B. Explicit expression for $h(w,\bar{w})$}
The integral over $\tau$ in formula \eqref{hf2} can be computed
(via partial integration). Below we provide the final result of
this calculation

\begin{multline}
h(w,\bar{w})=\int_0^1 dt\Bigg\{ \frac{1}{z^2}\int\frac{\dr \bar{u} \, \dr\bar{v}}{(2\pi)^2} e^{-\frac{i}{z}\bar{u}\bar{v}} f(\bar{w}+(1-t)w+\bar{u})f(\bar{w}+\bar{v}-tw)+\\
+ \frac{1}{z^2}  \int \frac{\dr u\, \dr v}{(2\pi)^2} e^{\frac{i}{z}uv}\, f(w+u+(1-t)\bar{w}) f(y+v-t\bar{y})+\\
+\frac{1}{2z^2} \left(w^\sigma \frac{\partial}{\partial w^\sigma}\right) \int \frac{\dr \bar{u}\, \dr \bar{v}}{(2\pi)^2} e^{-\frac{i}{z}\bar{u}\bar{v}}\, f(\bar{w}+(1-t)w+\bar{u}) f(\bar{w}+\bar{v}-tw)+\\
+\frac{1}{2 z^2} \left(\bar{w}^\sigma \frac{\partial}{\partial \bar{w}^\sigma}\right) \int \frac{\dr u\, \dr v}{(2\pi)^2} e^{\frac{i}{z}uv}\, f(w+u+(1-t)\bar{w}) f(w+v-t\bar{w})-\\
-\frac{2}{z^2}\int \frac{\dr u\, \dr v}{(2\pi)^2} e^{\frac{i}{z}u_\alpha v^\alpha} f(u)f(v)-\\
-\frac{1}{2}   f(tw)\left(w^\sigma \frac{\partial}{\partial w^\sigma}\right)\Big(f(\bar{w}+tw)+f(\bar{w}-tw)\Big)-\\
-\frac{1}{2}  f(t\bar{w})\left(\bar{w}^\sigma \frac{\partial}{\partial \bar{w}^\sigma}\right)\big(f(w+t\bar{w})-f(w-t\bar{w})\big)\Bigg\}+\mathbf{c}(z).
\end{multline}



\begin{thebibliography}{10}

\bibitem{Maldacena:1997re}
J.~M. Maldacena, ``{The Large N limit of superconformal field
theories and supergravity},'' {\em Adv. Theor. Math. Phys.},
vol.~2, pp.~231--252, 1998, hep-th/9711200.

\bibitem{Gubser:1998bc}
S.~S. Gubser, I.~R. Klebanov, and A.~M. Polyakov, ``{Gauge theory
correlators from noncritical string theory},'' {\em Phys. Lett.
B}, vol.~428, pp.~105--114, 1998, hep-th/9802109.

\bibitem{Witten:1998qj}
E.~Witten, ``{Anti-de Sitter space and holography},'' {\em Adv.
Theor. Math. Phys.}, vol.~2, pp.~253--291, 1998, hep-th/9802150.

\bibitem{Didenko:2021vui}
V.~E.~Didenko and A.~V.~Korybut,``Planar solutions of higher-spin
theory. Part I. Free field level,'' JHEP \textbf{08} (2021), 144
doi:10.1007/JHEP08(2021)144 [arXiv:2105.09021 [hep-th]].


\bibitem{arXiv:9808032}
D.~Birmingham, ``{Topological black holes in Anti-de Sitter space},'' {\em
  Class. Quant. Grav.}, vol.~16, pp.~1197--1205, 1999, hep-th/9808032.

\bibitem{Vasiliev:1999ba}
M.~A. Vasiliev, ``{Higher spin gauge theories: Star product and
AdS space},'' 10 1999, hep-th/9910096.

\bibitem{more}
M.~A. Vasiliev, ``{More on equations of motion for interacting
massless fields of all spins in (3+1)-dimensions},'' {\em Phys.
Lett. B}, vol.~285, pp.~225--234, 1992.


\bibitem{Gelfond:2018vmi}
O.~A. Gelfond and M.~A. Vasiliev, ``{Homotopy Operators and
Locality Theorems
  in Higher-Spin Equations},'' {\em Phys. Lett. B}, vol.~786, pp.~180--188,
  2018, 1805.11941.

\bibitem{Didenko:2018fgx}
V.~E. Didenko, O.~A. Gelfond, A.~V. Korybut, and M.~A. Vasiliev,
``{Homotopy
  Properties and Lower-Order Vertices in Higher-Spin Equations},'' {\em J.
  Phys. A}, vol.~51, no.~46, p.~465202, 2018, 1807.00001.


\bibitem{Aharony:2020omh}
O.~Aharony, S.~M. Chester, and E.~Y. Urbach, ``{A Derivation of AdS/CFT for
  Vector Models},'' {\em JHEP}, vol.~03, p.~208, 2021, 2011.06328.


\bibitem{Das:2003vw}
S.~R. Das and A.~Jevicki, ``{Large N collective fields and holography},'' {\em
  Phys. Rev. D}, vol.~68, p.~044011, 2003, hep-th/0304093.

\bibitem{Jevicki:2015sla}
A.~Jevicki and J.~Yoon,
JHEP \textbf{02} (2016), 090, 1503.08484

\bibitem{Prokushkin:1998bq}
S.~F.~Prokushkin and M.~A.~Vasiliev, ``Higher spin gauge
interactions for massive matter fields in 3-D AdS space-time,''
Nucl. Phys. B \textbf{545} (1999), 385
doi:10.1016/S0550-3213(98)00839-6 [arXiv:hep-th/9806236 [hep-th]].

\bibitem{Sezgin:2005pv}
E.~Sezgin and P.~Sundell, ``{An Exact solution of 4-D higher-spin
gauge
  theory},'' {\em Nucl. Phys. B}, vol.~762, pp.~1--37, 2007, hep-th/0508158.

\bibitem{Iazeolla:2007wt}
C.~Iazeolla, E.~Sezgin, and P.~Sundell, ``{Real forms of complex
higher spin
  field equations and new exact solutions},'' {\em Nucl. Phys. B}, vol.~791,
  pp.~231--264, 2008, 0706.2983.

\bibitem{Didenko:2009td}
V.~E. Didenko and M.~A. Vasiliev, ``{Static BPS black hole in 4d
higher-spin
  gauge theory},'' {\em Phys. Lett. B}, vol.~682, pp.~305--315, 2009,
  0906.3898.
\newblock [Erratum: Phys.Lett.B 722, 389 (2013)].

\bibitem{Iazeolla:2011cb}
C.~Iazeolla and P.~Sundell, ``{Families of exact solutions to
Vasiliev's 4D
  equations with spherical, cylindrical and biaxial symmetry},'' {\em JHEP},
  vol.~12, p.~084, 2011, 1107.1217.

\bibitem{Bourdier:2014lya}
J.~Bourdier and N.~Drukker, ``{On Classical Solutions of 4d
Supersymmetric
  Higher Spin Theory},'' {\em JHEP}, vol.~04, p.~097, 2015, 1411.7037.

\bibitem{Iazeolla:2015tca}
C.~Iazeolla and J.~Raeymaekers, ``{On big crunch solutions in
  Prokushkin-Vasiliev theory},'' {\em JHEP}, vol.~01, p.~177, 2016, 1510.08835.

\bibitem{Sundell:2016mxc}
P.~Sundell and Y.~Yin, ``{New classes of bi-axially symmetric
solutions to
  four-dimensional Vasiliev higher spin gravity},'' {\em JHEP}, vol.~01,
  p.~043, 2017, 1610.03449.

\bibitem{Iazeolla:2017vng}
C.~Iazeolla and P.~Sundell, ``{4D Higher Spin Black Holes with
Nonlinear Scalar
  Fluctuations},'' {\em JHEP}, vol.~10, p.~130, 2017, 1705.06713.

\bibitem{Iazeolla:2017dxc}
C.~Iazeolla, E.~Sezgin, and P.~Sundell, ``{On Exact Solutions and
Perturbative
  Schemes in Higher Spin Theory},'' {\em Universe}, vol.~4, no.~1, p.~5, 2018,
  1711.03550.

\bibitem{Aros:2017ror}
R.~Aros, C.~Iazeolla, J.~Nore\~na, E.~Sezgin, P.~Sundell, and
Y.~Yin, ``{FRW and domain walls in higher spin gravity},'' {\em
JHEP}, vol.~03, p.~153, 2018, 1712.02401.

\bibitem{Bekaert:2015tva}
X.~Bekaert, J.~Erdmenger, D.~Ponomarev, and C.~Sleight, ``{Quartic
AdS Interactions in Higher-Spin Gravity from Conformal Field
Theory},'' {\em JHEP}, vol.~11, p.~149, 2015, 1508.04292.

\bibitem{Sleight:2017pcz}
C.~Sleight and M.~Taronna, ``{Higher-Spin Gauge Theories and Bulk
Locality},'' {\em Phys. Rev. Lett.}, vol.~121, no.~17, p.~171604,
2018, 1704.07859.


\bibitem{Vasiliev:2016xui}
M.~A. Vasiliev, ``{Current Interactions and Holography from the
0-Form Sector of Nonlinear Higher-Spin Equations},'' {\em JHEP},
vol.~10, p.~111, 2017, 1605.02662.


\bibitem{Didenko:2019xzz}
V.~E. Didenko, O.~A. Gelfond, A.~V. Korybut, and M.~A. Vasiliev,
``{Limiting
  Shifted Homotopy in Higher-Spin Theory and Spin-Locality},'' {\em JHEP},
  vol.~12, p.~086, 2019, 1909.04876.

\bibitem{Gelfond:2019tac}
O.~A. Gelfond and M.~A. Vasiliev, ``{Spin-Locality of Higher-Spin
Theories and Star-Product Functional Classes},'' {\em JHEP},
vol.~03, p.~002, 2020, 1910.00487.

\bibitem{Didenko:2020bxd}
V.~E. Didenko, O.~A. Gelfond, A.~V. Korybut, and M.~A. Vasiliev,
``{Spin-locality of $\eta^{2}$ and $ {\overline{\eta}}^2 $ quartic
higher-spin vertices},'' {\em JHEP}, vol.~12, p.~184, 2020,
2009.02811.

\bibitem{David:2020ptn}
A.~David and Y.~Neiman, ``Higher-spin symmetry vs. boundary
locality, and a rehabilitation of dS/CFT,'' JHEP \textbf{10}
(2020), 127 doi:10.1007/JHEP10(2020)127 [arXiv:2006.15813
[hep-th]].

\bibitem{Gelfond:2021two}
O.~A.~Gelfond and A.~V.~Korybut,
``Manifest form of the spin-local higher-spin vertex $\varUpsilon ^{\eta \eta }_{\omega CCC}$,''
Eur. Phys. J. C \textbf{81} (2021) no.7, 605
doi:10.1140/epjc/s10052-021-09401-4
[arXiv:2101.01683 [hep-th]].


\bibitem{Didenko:2008va}
V.~E. Didenko, A.~S. Matveev, and M.~A. Vasiliev, ``{Unfolded
Description of AdS(4) Kerr Black Hole},'' {\em Phys. Lett. B},
vol.~665, pp.~284--293, 2008, 0801.2213.

\bibitem{Didenko:2009tc}
V.~E. Didenko, A.~S. Matveev, and M.~A. Vasiliev, ``{Unfolded
Dynamics and Parameter Flow of Generic AdS(4) Black Hole},'' 1
2009, 0901.2172.

\bibitem{Didenko:2011ir}
V.~E.~Didenko, ``Coordinate independent approach to 5d black
holes,'' Class. Quant. Grav. \textbf{29} (2012), 025009
doi:10.1088/0264-9381/29/2/025009 [arXiv:1108.4321 [hep-th]].

\bibitem{Monteiro:2014cda}
R.~Monteiro, D.~O'Connell and C.~D.~White, ``Black holes and the
double copy,'' JHEP \textbf{12} (2014), 056
doi:10.1007/JHEP12(2014)056 [arXiv:1410.0239 [hep-th]].

\bibitem{Luna:2018dpt}
A.~Luna, R.~Monteiro, I.~Nicholson and D.~O'Connell, ``Type D
Spacetimes and the Weyl Double Copy,'' Class. Quant. Grav.
\textbf{36} (2019), 065003 doi:10.1088/1361-6382/ab03e6
[arXiv:1810.08183 [hep-th]].

\bibitem{Kawai:1985xq}
H.~Kawai, D.~C.~Lewellen and S.~H.~H.~Tye,
Nucl. Phys. B \textbf{269} (1986), 1-23
doi:10.1016/0550-3213(86)90362-7

\bibitem{Adamo:2020qru}
T.~Adamo and A.~Ilderton, ``Classical and quantum double copy of
back-reaction,'' JHEP \textbf{09} (2020), 200
doi:10.1007/JHEP09(2020)200 [arXiv:2005.05807 [hep-th]].

\bibitem{Godazgar:2020zbv}
H.~Godazgar, M.~Godazgar, R.~Monteiro, D.~P.~Veiga and C.~N.~Pope,
``Weyl Double Copy for Gravitational Waves,'' Phys. Rev. Lett.
\textbf{126} (2021) no.10, 101103
doi:10.1103/PhysRevLett.126.101103 [arXiv:2010.02925 [hep-th]].

\bibitem{Ferrero:2020vww}
P.~Ferrero and D.~Francia, ``On the Lagrangian formulation of the
double copy to cubic order,'' JHEP \textbf{02} (2021), 213
doi:10.1007/JHEP02(2021)213 [arXiv:2012.00713 [hep-th]].

\bibitem{White:2020sfn}
C.~D.~White, ``Twistorial Foundation for the Classical Double
Copy,'' Phys. Rev. Lett. \textbf{126} (2021) no.6, 061602
doi:10.1103/PhysRevLett.126.061602 [arXiv:2012.02479 [hep-th]].

\bibitem{Borsten:2021hua}
L.~Borsten, B.~Jur\v{c}o, H.~Kim, T.~Macrelli, C.~Saemann and
M.~Wolf, ``Double Copy from Homotopy Algebras,'' Fortsch. Phys.
\textbf{69} (2021), 2100075 doi:10.1002/prop.202100075
[arXiv:2102.11390 [hep-th]].

\bibitem{Chacon:2021wbr}
E.~Chac\'on, S.~Nagy and C.~D.~White, ``The Weyl double copy from
twistor space,'' JHEP \textbf{05} (2021), 2239
doi:10.1007/JHEP05(2021)239 [arXiv:2103.16441 [hep-th]].

\bibitem{Zhou:2021gnu}
X.~Zhou, ``Double Copy Relation in AdS Space,'' Phys. Rev. Lett.
\textbf{127} (2021) no.14, 141601
doi:10.1103/PhysRevLett.127.141601 [arXiv:2106.07651 [hep-th]].

\bibitem{Chacon:2021hfe}
E.~Chac\'on, A.~Luna and C.~D.~White, ``The double copy of the
multipole expansion,'' [arXiv:2108.07702 [hep-th]].

\bibitem{Godazgar:2021iae}
H.~Godazgar, M.~Godazgar, R.~Monteiro, D.~P.~Veiga and C.~N.~Pope,
``Asymptotic Weyl Double Copy,'' [arXiv:2109.07866 [hep-th]].

\bibitem{Adamo:2021dfg}
T.~Adamo and U.~Kol, ``Classical double copy at null infinity,''
[arXiv:2109.07832 [hep-th]].

\bibitem{Vasiliev:2012vf}
M.~A. Vasiliev, ``{Holography, Unfolding and Higher-Spin
Theory},'' {\em J. Phys. A}, vol.~46, p.~214013, 2013, 1203.5554.

\bibitem{Didenko:2017lsn}
V.~E. Didenko and M.~A. Vasiliev, ``{Test of the local form of
higher-spin equations via AdS / CFT},'' {\em Phys. Lett. B},
vol.~775, pp.~352--360, 2017, 1705.03440.



\bibitem{Vasiliev:1988sa}
M.~A. Vasiliev, ``{Consistent Equations for Interacting Massless
Fields of All Spins in the First Order in Curvatures},'' {\em
Annals Phys.}, vol.~190,  pp.~59--106, 1989.

\bibitem{Sezgin:2003pt}
E.~Sezgin and P.~Sundell, ``Holography in 4D (super) higher spin
theories and a test via cubic scalar couplings,'' JHEP \textbf{07}
(2005), 044 doi:10.1088/1126-6708/2005/07/044
[arXiv:hep-th/0305040 [hep-th]].

\bibitem{Didenko:2015pjo}
V.~E.~Didenko, N.~G.~Misuna and M.~A.~Vasiliev, ``Charges in
nonlinear higher-spin theory,'' JHEP \textbf{03} (2017), 164
doi:10.1007/JHEP03(2017)164 [arXiv:1512.07626 [hep-th]].

\bibitem{Sleight:2016dba}
C.~Sleight and M.~Taronna, ``{Higher Spin Interactions from Conformal Field
  Theory: The Complete Cubic Couplings},'' {\em Phys. Rev. Lett.}, vol.~116,
  no.~18, p.~181602, 2016, 1603.00022.

\bibitem{GY1}
S.~Giombi and X.~Yin, ``{Higher Spin Gauge Theory and Holography: The
  Three-Point Functions},'' {\em JHEP}, vol.~09, p.~115, 2010, 0912.3462.

\bibitem{Didenko:2012vh}
V.~E. Didenko and E.~D. Skvortsov, ``{Towards higher-spin holography in ambient
  space of any dimension},'' {\em J. Phys. A}, vol.~46, p.~214010, 2013,
  1207.6786.

\bibitem{Gelfond:2015poa}
O.~A.~Gelfond and M.~A.~Vasiliev, ``Symmetries of higher-spin
current interactions in four dimensions,'' Theor. Math. Phys.
\textbf{187} (2016) no.3, 797-812 doi:10.1134/S0040577916060015
[arXiv:1510.03488 [hep-th]].

\bibitem{Bolotin:1999fa}
K.~I. Bolotin and M.~A. Vasiliev, ``{Star product and massless
free field dynamics in AdS(4)},'' {\em Phys. Lett. B}, vol.~479,
pp.~421--428, 2000, hep-th/0001031.

\bibitem{Nagaraj:2019zmk}
B.~Nagaraj and D.~Ponomarev, ``{Spinor-helicity formalism for
massless fields in AdS$_{4}$. Part II. Potentials},'' {\em JHEP},
vol.~06, p.~068, 2020, 1912.07494.

\bibitem{Metsaev:1999kb}
R.~R.~Metsaev, ``IIB supergravity and various aspects of light
cone formalism in AdS space-time,'' [arXiv:hep-th/0002008
[hep-th]].

\bibitem{Barvinsky:2015wvz}
A.~O.~Barvinsky, ``CFT driven cosmology and conformal higher spin
fields,'' Phys. Rev. D \textbf{93} (2016) no.10, 103530
doi:10.1103/PhysRevD.93.103530 [arXiv:1511.07625 [hep-th]].

\end{thebibliography}


\end{document}